\documentclass{article}

\usepackage[preprint]{neurips_data_2024}

\usepackage[utf8]{inputenc} 
\usepackage[T1]{fontenc}    
\usepackage{hyperref}       
\usepackage{url}            
\usepackage{booktabs}       
\usepackage{amsfonts}       
\usepackage{nicefrac}       
\usepackage{microtype}      
\usepackage{xcolor}         

\usepackage{multirow}
\usepackage{graphicx}
\usepackage{subcaption}
\usepackage{wrapfig}
\usepackage{listings}
\input{Definitions}

\title{Flow-Bench: A Dataset for Computational Workflow Anomaly Detection}

\author{%
  George Papadimitriou\footnotemark[2]$^{\ }$ \thanks{Equal contribution.}
  \And 
  Hongwei Jin\footnotemark[3] $\ ^*$
  \And
  Cong Wang\footnotemark[4] 
  \And
  Rajiv Mayani\footnotemark[2] \\
  \And
  Krishnan Raghavan\footnotemark[3]
  \And
  Anirban Mandal\footnotemark[4]
  \And
  Prasanna Balaprakash\footnotemark[5]
  \And
  Ewa Deelman\footnotemark[2]  \\
  \and
  \\
  \footnotemark[2]$\ $
  University of Southern California \\
  Los Angeles, CA \\
  \texttt{\{georgpap,mayani,deelman\}@isi.edu} \\
  \and
  \footnotemark[3]$\ $
  Argonne National Laboratory\\
  Lemont, IL\\
  \texttt{\{jinh, kraghavan\}@anl.gov} \\
  \and
  \footnotemark[4]$\ $
  RENCI \\
  Chapel Hill, NC \\
  \texttt{\{cwang, anirban\}@renci.org} \\
  \and
  \footnotemark[5]$\ $
  Oak Ridge National Laboratory \\
  Oak Ridge, TN \\
  \texttt{pbalapra@ornl.gov} \\
}

\begin{document}

\maketitle

\begin{abstract}
  A computational workflow, also known as workflow, consists of tasks that must be executed in a specific order to attain a specific goal. Often, in fields such as biology, chemistry, physics, and data science, among others, these workflows are complex and are executed in large-scale, distributed, and heterogeneous computing environments prone to failures and performance degradation. Therefore, anomaly detection for workflows is an important paradigm that aims to identify unexpected behavior or errors in workflow execution. This crucial task to improve the reliability of workflow executions can be further assisted by machine learning-based techniques. However, such application is limited, in large part, due to the lack of open datasets and benchmarking. To address this gap, we make the following contributions in this paper: (1) we systematically inject anomalies and collect raw execution logs from workflows executing on distributed infrastructures; (2) we summarize the statistics of new datasets, and provide insightful analyses; (3) we convert workflows into tabular, graph and text data, and benchmark with supervised and unsupervised anomaly detection techniques correspondingly. The presented dataset and benchmarks allow examining the effectiveness and efficiency of scientific computational workflows and identifying potential research opportunities for improvement and generalization. The dataset and benchmark code are publicly available \url{https://poseidon-workflows.github.io/FlowBench/} under the MIT License.
\end{abstract}

\section{Introduction}
\label{sec:intro}

Computational workflows that run on distributed and parallel computing environments, such as Southern California Earthquake Center's (SCEC) Cybershake~\citep{callaghan2014} and LIGO's PyCBC~\citep{pycbc-software}, have become an integral part of scientific, engineering, and data science research, enabling the simulation and modeling of complex otherwise impossible complex phenomena. However, in recent years, the scale and complexity of computational workflow systems have increased significantly. This increase has made these workflow applications prone to various types of hardware and software faults, including performance anomalies, crashes, and security breaches. These faults can cause significant disruptions to high-performance computing (HPC) applications, resulting in lost productivity, wasted resources, and potentially serious consequences for scientific discoveries.

To address this challenge, anomaly detection has emerged as a promising approach for detecting and diagnosing workflow failures. Anomaly detection refers to the process of identifying patterns or behaviors that deviate significantly from expected~(or normal) behavior. Detection of anomalies enables system administrators and users to take proactive measures to prevent or mitigate the impact of faults, improving the overall reliability and efficiency of the workflow. Recently, various research efforts have been devoted to developing and evaluating anomaly detection techniques for computational workflows. While anomaly detection in workflows is very desirable, key challenges still persist: (1) there is no template for determining the relevance of deviations in the workflow settings, (2) most of the time, the training data is comprised of several orders of magnitude more data points corresponding to the normal than the anomaly, and (3) more often than not, labeling the anomalies in workflows incurs of lots of time and effort.

These fundamental challenges make the anomaly detection problem hard. Although standard anomaly detection benchmarks such as \cite{zhao2019pyod} and \cite{liu2022bond} can be applied to workflows. As is shown in this work, these methods are not very effective and do not scale well with the size of the workflow as well. Specifically, we show in Appendix~\ref{appx:benchmark_results} that many readily available anomaly detection methods result in out-of-memory (OOM) errors when encountering simple workflow data. On the other hand, the methods that do manage to work on the workflow data do not provide desirable precision. Moreover, there is still a lack of consensus on the best practices and benchmark datasets that facilitate the evaluation of different anomaly detection techniques on computational workflows, making it difficult to compare and reproduce the results of different studies.

These challenges indicate the need for careful analysis and study of workflow data, but the benchmarks needed for such analysis of workflows are not available in the literature. In this work, we present a set of novel workflows along with best practices and preliminary analysis that will provide the tools for future research to test and develop tools to facilitate anomaly detection on workflows. In particular, we present a benchmark study of anomaly detection techniques for workflows. We introduce a set of new workflow data that we have collected for anomaly detection and evaluate the performance of several state-of-the-art anomaly detection techniques, including tabular data anomaly detection~\citep{zhao2019pyod}, graph data anomaly detection~\citep{liu2022bond, kipf2017semi}, and text data in supervise fine-tuned language models~\citep{guo2021logbert} in supervised and unsupervised way. Our evaluation is based on both supervised and unsupervised learning with several performance metrics, including accuracy, ROC-AUC score, F1-score, top-k score, etc. We compare the strengths and weaknesses of different anomaly detection techniques and reveal insights into their applicability and limitations in scientific workflow scenarios.

In spite of our benchmark being focused on computational workflows, it is important to note that challenges such as non-stationarity in input data distribution, differences in the number of nodes and features, the size of the data, and especially the lack of labels, provide a real-world example that can be readily utilized for the development of anomaly detection tools for the larger machine learning~(ML) community. In contrast, popular benchmark tools such as PyOD~\citep{zhao2019pyod} and PyGOD~\citep{liu2022bond} do not provide these unique features.

\vspace{-1em}
\subsection{Contributions}
\label{subsec:contributions}

We introduce a set of workflow data that allows the careful study and development of novel anomaly detection approaches for computational workflows. We provide a comprehensive study to demonstrate the utility of our data with already available benchmarks and why novel methods are necessary.

\section{Related Works}
\label{sec:related}

There are several techniques that have been proposed for anomaly detection. Statistical methods, machine learning~(ML) methods, rule-based methods, and hybrid methods are the four major categories of these methods.  Statistical methods formulate the anomaly detection problem as the detection of outliers in a distribution, and therefore, methods such as clustering~\citep{sohn2023anomaly}, principal component analysis (PCA)~\citep{ma2023mppca}, and auto-regressive integrated moving average (ARIMA) models~\citep{goldstein2023special} are commonly utilized. On the other hand, the ML methods involve learning a classifier map between the anomaly class and the input data. These methods have been developed for both tabular and graph data, and popular methods include (graph) neural networks~\citep{yuan2021higher}, decision trees~\citep{breiman2001random}, and support vector machines (SVMs)~\citep{scholkopf2001estimating}. Hybrid methods take the probabilistic efficiency of statistical methods and combine it with the ability of ML methods to characterize complex distribution to improve the accuracy and robustness of anomaly detection~\citep{kriegel2008angle}. Most recently, leveraging supervised fine-tuning and training large language models~(LLMs)~\citep{vaswani2017attention} on anomaly-labeled text data enables them to amplify their ability to detect unusual language patterns specific to the domain, making them highly effective for anomaly detection.

Another class of anomaly detection methods involves rule-based methods or expert systems where a set of rules must be defined to characterize normal system behavior. These rules can then be used to detect anomalies~\citep{liu2008isolation} in the data. These rules can either be based on expert knowledge or derived from historical data or ML or probabilistic models that encode normal information.
While a non-exhaustive list of anomaly detection methods is presented above, all these methods can be potentially used in workflows. However, it is important to note that the choice of technique(s) for anomaly detection in workflows depends on the specific requirements and characteristics of the system being monitored, as well as the goals of the anomaly detection. For instance, a graph neural network (GNN)--based approach is more suitable for workflows than methods designed for tabular data because the dependency structure present in workflow data is rather important to characterize anomalies in computational workflows. Furthermore, computational workflows are unique in the setup as each workflow differs in the number of jobs, the dependency graph, and other structures. While training a separate anomaly detection approach for each workflow is inefficient, typical approaches discussed above, except the graph neural network-based, are unsuitable when the input data dimensions change for every computational workflow. On the other hand, workflow data can be rather large, and complex ML-based approaches run out of memory when trying to detect anomalies in these datasets.

In the past, datasets containing system logs of large systems (e.g., BlueGene, Thunderbird) have been released to the community~\cite{oliner2007dsn}. However, these logs are unstructured, and they don't match the statistics and events to the workloads that were executed on these systems. Providing a dataset directly correlating system logs and the workflow application is important in creating new ML methods that better capture and identify workflow anomalies.

\section{Dataset}
\label{sec:datasets}

The datasets used for benchmarking anomaly detection in computational workflows should represent the data types encountered in real applications. The datasets should include both normal and anomalous data to evaluate the performance of the base method. Some important statistics that should be considered when selecting datasets are:
\begin{itemize}
  \item \textbf{Data size}: The dataset should be large enough to capture the variability in the data but not so large that it becomes computationally impractical to analyze and benchmark.
  \item \textbf{Data dimensionality}: The dataset's number of features or variables should represent the data types encountered in real applications.
  \item \textbf{Data distribution}: The dataset should represent the typical data distribution encountered in real applications.
\end{itemize}

\subsection{Workflow Applications}
Towards this end, we choose twelve workflow applications that are unique in their characteristics and are representative of real-world applications across multiple domains.

\vspace{-1em}
\paragraph{1000 Genome Workflow}
The 1000 genome project~\citep{1000genome-project} provides a reference for human variation, having reconstructed the genomes of 2,504 individuals across 26 different populations. The test case we have here identifies mutational overlaps using data from the 1000 genomes project in order to provide a null distribution for rigorous statistical evaluation of potential disease-related mutations. This workflow (Figure~\ref{fig:1000genome} in Appendix~\ref{appx:workflow_diagrams}) is composed of five different tasks.

\vspace{-1em}
\paragraph{Montage Workflow}
Montage is an astronomical image toolkit~\citep{2010ascl.soft10036J} with components for re-projection, background matching, co-addition, and visualization of FITS files. Montage workflows typically follow a predictable structure based on the inputs, with each stage of the workflow often taking place in discrete levels separated by some synchronization/reduction tasks (mConcatFit and mBgModel). The workflow (Figure~\ref{fig:montage} in Appendix~\ref{appx:workflow_diagrams}) uses the Montage to transform astronomical images into custom mosaics, and is composed of eight different tasks.
\vspace{-1em}
\paragraph{Predict Future Sales~(Data Science) Workflow}
The predict future sales workflow~\citep{kaggle-predict-future-sales} tackles Kaggle's predict future sales competition by receiving daily historical sales data from January 2013 to October 2015 and attempting to predict the sales for November 2015. It includes preprocessing, feature engineering, dataset segmentation based on type and sales performance, hyperparameter tuning, training of 3 XGBoost models for each group, and an ensemble technique to combine the model predictions into a single output file. Figure~\ref{fig:predict-future-sales} in Appendix~\ref{appx:workflow_diagrams} illustrates the eleven-task workflow.


\vspace{-1em}
\paragraph{Variant Calling Workflow}
The variant calling workflow contains all the steps to detect variations and identify changes in a DNA sequence compared to a reference genome. This workflow is adapted from the Data Carpentry Lesson - Data Wrangling and Processing for Genomics~\citep{datacarpentry-variant-calling}, that provides a pipeline to align population samples to the reference genome and detects the differences that exist. Variant calling is a very important process in bioinformatics that scientists have to perform. Figure~\ref{fig:variant-calling} in Appendix~\ref{appx:workflow_diagrams} presents the workflow, which is composed of six different tasks.

\vspace{-1em}
\paragraph{CASA Workflows}
The Collaborative Adaptive Sensing of the Atmosphere (CASA)~\citep{2009:BAMS:CASA} has deployed a network of Doppler radars in the Dallas-Fort Worth (DFW) area, in order to improve observations, predictions and responses to hazardous weather events.\\
The \textit{Wind Speed} workflow~\citep{lyons-escience-2019} is one of CASA's production pipelines that ingests single radar base data from all of the radars in the network and creates a gridded product representing the maximum observed wind speeds in the area. The workflow identifies areas of severe wind and correlates them with the location of critical infrastructure, such as hospitals. In case critical infrastructure is impacted, the workflow dispatches email alerts to the impacted locations. The workflow (Figure~\ref{fig:casa-wind} in Appendix~\ref{appx:workflow_diagrams}) calculates the maximum observed wind velocity across multiple minutes of radar data, and has five different tasks.\\
The \textit{Nowcast} workflow~\citep{lyons-escience-2019} computes short-term advection forecasts, by splitting gridded reflectivity radar data into 31 grids and computing reflectivity predictions over the next 30 minutes. Every minute new predictions need to be generated and Figure~\ref{fig:casa-nowcast} in Appendix~\ref{appx:workflow_diagrams} presents a nowcast workflow that processes data across multiple minutes using three different tasks.


\vspace{-1em}
\paragraph{Soil Moisture Spatial Inference Engine Workflow}
The soil moisture spatial inference engine (SOMOSPIE) workflow processes spatial environmental data to generate training and evaluation datasets and produces machine learning models that can predict the moisture of the soil in the future. SOMOSPIE is a modular workflow with multiple stages and Figure~\ref{fig:somospie} in Appendix~\ref{appx:workflow_diagrams} presents the data preparation part of the workflow to produce the training and evaluation data sets. The workflow is consisted of eight different tasks.


\vspace{-1em}
\paragraph{PyCBC Workflows}
PyCBC~\citep{pycbc-software} is a software package used to explore astrophysical sources of gravitational waves. It contains algorithms to analyze gravitational-wave data, detect coalescing compact binaries, and make bayesian inferences from gravitational-wave data collected by detectors (e.g., LIGO~\citep{Aasi_2015}).\\
The \textit{PyCBC Inference}~\citep{pycbc-inference} workflow performs parameter estimation on gravitational-wave signals using Bayesian inference methods to incorporate prior knowledge and quantify uncertainties in the parameter estimates. Figure~\ref{fig:pycbc-inference} in Appendix~\ref{appx:workflow_diagrams} presents the workflow that's consisted of nine different tasks.\\
The \textit{PyCBC Search}~\citep{pycbc-search} workflow sifts through noisy detector data to identify gravitational waves. It's a complex workflow application that dynamically steers itself using at-runtine generated subworkflows with over fifty different tasks (Figure~\ref{fig:pycbc-search} in Appendix~\ref{appx:workflow_diagrams}).



\vspace{-1em}
\paragraph{EHT Workflows}
The Event Horizon Telescope (EHT) workflows reproduce the first image of a black hole in the galaxy M87 and there are three image processing pipelines~\citep{patel-eht}.
The \textit{EHT Difmap} workflow performs image reconstruction by employing iterative convolution with difference mapping. Figure~\ref{fig:eht-difmap} in Appendix~\ref{appx:workflow_diagrams} shows the workflow with four different tasks.\\
The \textit{EHT Imaging} workflow performs image reconstruction using the regularized maximum likelihood (RML) approach. Figure~\ref{fig:eht-imaging} in Appendix~\ref{appx:workflow_diagrams} shows the workflow with two different tasks.\\
The \textit{EHT Smili} workflow uses the Sparse Modeling Imaging Library for Interferometry (SMILI) to introduce pre-calibrated datasets and perform image reconstruction using the RML approach. Figure~\ref{fig:eht-smili} in Appendix~\ref{appx:workflow_diagrams} shows the workflow with three different tasks.

\subsection{Data Collection}
\label{subsec:data_collection}

To execute the workflows described above, manage their execution and collect events and statistics, we used the Pegasus Workflow Management System (WMS)~\citep{deelman-fgcs-2015}.

\vspace{-.5em}
\subsubsection{Workflow Model}
In Pegasus, workflows are represented as Directed Acyclic Graphs (DAGs), where nodes represent jobs, and edges represent sequential dependencies and data dependencies between the jobs. A job can only be submitted when all its dependencies have been met.
For a more formal definition, a workflow is described as a DAG $\mathcal{G}=(\mathcal{V},\mathcal{E})$, where $\mathcal{V} = \{ v_1, \cdots, v_n \}$ represents the set of $n$ jobs and $\mathcal{E} \subseteq \mathcal{V}^2$ represents data dependencies between jobs. If $e_{ij} = (v_i , v_j) \in \mathcal{E}$, job $v_i$ must complete its execution before $v_j$ can start, i.e., a directed edge $v_i \rightarrow v_j$.
We define $succ(v_i)= \{v_k \,|\, (v_i, v_k) \in \mathcal{E}\} $ (resp. $pred(v_i) = \{v_k \,|\, (v_k, v_i) \in \mathcal{E}\})$ to be the successors (resp. predecessors) of the job $v_i \in \mathcal{V}$. A job ($v_i$) of a Pegasus workflow can be classified into one of these three categories: (1) \textit{Compute} jobs describe computational tasks; (2) \textit{Transfer} jobs move data to and from an execution site/node; (3) \textit{Auxiliary} jobs create working directories, clean up unused data or perform other Pegasus internal bookkeeping tasks.\\
Additionally, Pegasus collects provenance data and events during the execution of a workflow and associates them with the jobs that produced them. The Pegasus Panorama extensions~\citep{deelman2017panorama, pegasus-panorama,papadimitriou2021end} enhance the workflow DAG with features containing execution metrics and infrastructure statistics per job ($v_i$). These extensions enable the collection of execution traces of computational tasks, the collection of statistics of individual data transfers, the collection of infrastructure-related metrics, and all of them are stored in an Elasticsearch instance~\citep{elastic}.

\vspace{-.5em}
\subsubsection{Execution Infrastructure}
\label{susubsec:infrastructure}




To facilitate the workflow execution, we provisioned resources at the NSF Chameleon Cloud~\citep{keahey2020lessons} and the FABRIC testbed~\citep{fabric-baldin-2019} (Figure~\ref{fig:poseidon-deployment} in Appendix~\ref{appx:workflow_execution_infra}). We provisioned 4 VMs (16 Cores and 32GB RAM) on FABRIC and 3 Cascade Lake bare-metal nodes (48 Cores and 192GB RAM) on Chameleon. In FABRIC 1 had the role of \textit{submit node}, 1 had the role of \textit{data node} and 2 were responsible for network QoS to the workers (StarLight location). In Chameleon all nodes were \textit{Docker container executor nodes}~\citep{docker} located in the same region (University of Chicago - UC). The connectivity between the two testbeds/regions was established over a high-speed layer 2 VLAN (25 Gbps).
On our \textit{data node} we configured an Apache HTTP server, and on our \textit{submit node} we installed and configured the Pegasus WMS~\citep{deelman-fgcs-2015} binaries to manage the workflow execution, alongside with HTCondor~\citep{htcondor} to manage the baremetal resources we provisioned. To create an HTCondor pool of resources, we spawned HTCondor's Executor containers on the Docker nodes, and in the case of the Predict Future Sales workflow we created 18 containers with 8 cores and 32GB RAM each, and in all other cases we created 20 containers with 4 cores and 16GB RAM each. For the Predict Future Sales workflow, we assigned more resources to each executor container to accommodate the hyper-parameter optimization phase, which leverages multiple cores and demands more memory. Core and RAM resources were mapped to the containers using Docker's runtime options (cpuset, memory). To deliver data to the executor containers, we used the \textit{data node's} HTTP server, and we capped the download and upload speed to 1Gbps per HTCondor Executor container.

\vspace{-1em}
\subsubsection{Synthetic Anomalies}
\label{subsubsec:sythentic-anomalies-method}

To introduce synthetic anomalies, we used Docker's runtime options to limit and shape the performance of the spawned container executor nodes. Through the runtime options, we were able to configure the amount of CPU time an executor container is allotted and limit the average I/O each executor container can perform. These capabilities are supported via Linux's control groups version 2~\citep{cgroupsv2}. This approach offers sufficient isolation among the executors and allows us to obtain reproducible results from each type of experiment.

\vspace{-1em}
\subsection{Dataset Description}
\label{subsubsec:dataset-desc}

\begin{table}[t]
  \centering
  \caption{
  Overall Dataset Statistics. We label the nodes as Normal, CPU $K$ and HDD $K$.\\
  {\scriptsize Normal: No anomaly is introduced -- normal conditions.\\
  CPU $K$: M cores are advertised, but on some nodes, $K$ cores are not allowed to be used. ($K=2,3,4 ~ M=4,8 ~ and ~ K < M$)\\
  HDD $K$: On some executor nodes, the average write speed to the disk is capped at $K$ MB/s and the read speed at $(2\times K)$ MB/s. ($K=5, 10$)\\
  }}
  {
  \label{tab:dataset}
  \resizebox{\linewidth}{!}{
    \begin{tabular}{l||cc|c|ccc|cc|c|ccc|cc}
      \toprule
                                & \multicolumn{2}{c|}{DAG Information}
                                & \multicolumn{6}{c|}{\#DAG Executions}
                                & \multicolumn{6}{c}{\#Total Nodes per Type}                                                                              \\
      \cmidrule{2-15}
      \multirow{2}{*}{Workflow} & \multirow{2}{*}{Nodes}                     & \multirow{2}{*}{Edges}
                                & \multirow{2}{*}{Normal}                    & \multicolumn{3}{c|}{CPU} & \multicolumn{2}{c|}{HDD}
                                & \multirow{2}{*}{Normal}                    & \multicolumn{3}{c|}{CPU} & \multicolumn{2}{c}{HDD}                         \\
                                &                                            &
                                &                                            & 2                        & 3                        & 4    & 5     & 10
                                &                                            & 2                        & 3                        & 4    & 5     & 10    \\
      \midrule
      1000 Genome               & 137                                        & 289
                                & 50                                         & 100                      & 25                       & -    & 100   & 75
                                & 32261                                      & 5173                     & 756                      & -    & 5392  & 4368  \\
      Montage                   & 539                                        & 2838
                                & 51                                         & 46                       & 80                       & -    & 67    & 76
                                & 137229                                     & 4094                     & 11161                    & -    & 8947  & 11049 \\
      Predict Future Sales      & 165                                        & 581
                                & 100                                        & 88                       & 88                       & 88   & 88    & 88
                                & 72609                                      & 3361                     & 3323                     & 3193 & 3321  & 3293  \\
      Variant Calling           & 371                                        & 792
                                & 80                                         & 80                       & 80                       & -    & 75    & 80
                                & 115588                                     & 8287                     & 7222                     & -    & 7365  & 8083  \\
      CASA Wind Speed           & 162                                        & 342
                                & 150                                        & 200                      & 200                      & -    & 200   & 160
                                & 116836                                     & 8793                     & 8382                     & -    & 8305  & 5104  \\
      CASA Nowcast              & 2081                                       & 4029
                                & 101                                        & 80                       & 78                       & -    & 79    & 83
                                & 685045                                     & 49960                    & 46664                    & -    & 46104 & 48328 \\
      Soil Moisture             & 60                                         & 185
                                & 125                                        & 98                       & 97                       & -    & 92    & 93
                                & 24408                                      & 1706                     & 1428                     & -    & 1344  & 1414  \\
      PyCBC Inference           & 17                                         & 26
                                & 206                                        & 89                       & 74                       & -    & 67    & 66
                                & 6970                                       & 549                      & 326                      & -    & 388   & 301   \\
      PyCBC Search              & 368                                        & 1025
                                & 102                                        & 100                      & 100                      & -    & 104   & 100
                                & 151004                                     & 9495                     & 9039                     & -    & 8324  & 8346  \\
      EHT Difmap                & 33                                         & 59
                                & 142                                        & 93                       & 88                       & -    & 89    & 89
                                & 13000                                      & 1059                     & 737                      & -    & 877   & 860   \\
      EHT Imaging               & 12                                         & 18
                                & 212                                        & 76                       & 74                       & -    & 70    & 69
                                & 4908                                       & 354                      & 241                      & -    & 261   & 248   \\
      EHT Smili                 & 16                                         & 26
                                & 148                                        & 87                       & 84                       & -    & 90    & 93
                                & 6471                                       & 437                      & 325                      & -    & 382   & 417   \\
      \bottomrule
    \end{tabular}
  }}
\end{table}

Table~\ref{tab:dataset} presents statistics of the executable workflow DAGs on the provisioned resources. Throughout the data harvesting phase, we maintained the DAGs of the workflows unchanged, and the DAG information presented in Table~\ref{tab:dataset} reflects the number of nodes and edges of the final executable workflow DAGs submitted by Pegasus. The dataset contains 6352 workflow executions, under normal and 2 anomalous conditions, with different anomaly levels. By using the method of Section~\ref{subsubsec:sythentic-anomalies-method} to inject synthetic anomalies, the anomalies are introduced on a per DAG node level, and as a result, each DAG execution can have a different set of anomalous nodes, since HTCondor negotiates which worker will accept a job at submission time. The dataset contains 6 tags \textit{normal}, \textit{cpu\_2}, \textit{cpu\_3}, \textit{cpu\_4}, \textit{hdd\_5} and \textit{hdd\_10}, that label the DAG nodes appropriately. Finally, the dataset provides the raw logs generated by Pegasus WMS and our experimental harness, and a parsed version, where we combined the raw logs into a single log file for each DAG execution.
In the raw data we are offering (1) the workflows and the configurations used to execute them; (2) the original workflow submit directories of all DAG executions; (3) the workflow management system logs; (4) an experiment log file mapping the workflow DAG executions to the injected anomalies; (5) provenance data of each job, gathered by the WMS; (6) an Elasticsearch instance with all the captured workflow events, transfer events and resource utilization traces. Since we acquired the logs and metrics using an ephemeral private cloud instance, there is no sensitive information included in the dataset. The total size of the raw logs is over 200GB.


\subsubsection{Parsed Data}
\label{subsubsec:parsed_data}


\begin{wrapfigure}{r}{0.5\linewidth}
  \includegraphics[width=0.5\textwidth]{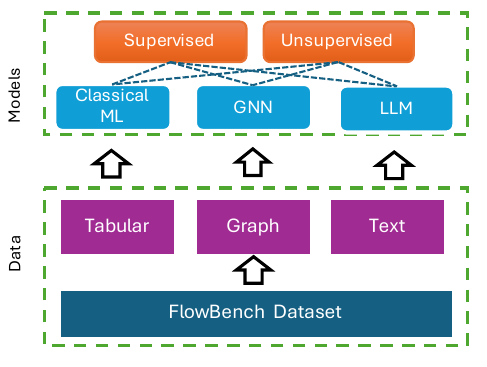}
  \vspace{-1em}
  \caption{Overview of FlowBench}
  \label{fig:flowbench-diagram}
\end{wrapfigure}

To generate the parsed version of the dataset, we combined information coming from the executable DAGs, the experiment log containing the anomaly labels, Pegasus' logs, and the events hosted in Elasticsearch (Figure~\ref{fig:raw-data-parsing}). During parsing, a CSV log file gets generated for every workflow DAG execution and is saved under the corresponding anomaly folder. The features present in the parsed data are listed in Table~\ref{tab:parsed-features} in the appendix with their corresponding description, and the total size of the parsed data is 783MB. More specific to text data, we prepare sentences to describe the job given the feature and values in the template of

``\texttt{FEAT1 is VAL1 FEAT2 is VAL2 ... FEATn is VALn, LABEL}''

which can be used in the supervised fine-tuning of LLMs. All of our assets will be released under the MIT License.

\vspace{-1em}
\section{ML Benchmarks on Computational Workflows}
\vspace{-1em}
\label{sec:bench_mark_design}
Anomaly detection in computational workflows is essential to ensure that experimental results are reliable and reproducible. A benchmark for anomaly detection in computational workflows would require selecting appropriate baseline methods and various metrics to evaluate the performance. We employ both the supervised and unsupervised approaches for the baselines due to the availability of labels in our scientific workflows.
Table~\ref{tab:algorithms} in appendix~\ref{appx:algorithms} provides a set of selected algorithms with their details.

\vspace{-1em}
\paragraph{Supervised learning.} Supervised anomaly detection models are trained on labeled data, where the job anomalies are already known. These models learn to identify anomalies based on the labeled data and can be used to detect anomalies in new, unlabeled data.
Exampled classical methods are:
support vector machine (SVM) \citep{hearst1998support}
, decision trees \citep{quinlan1986induction}
, random forest \citep{breiman2001random}
, etc. Specific to the DAGs, there are (semi)supervised learning algorithms that can learn the structural and feature information simultaneously, such as GraphSAGE \citep{hamilton2017inductive}, GAT \citep{velivckovic2017graph}, etc.

\vspace{-1em}
\paragraph{Unsupervised learning.}  Unsupervised anomaly detection models do not rely on labeled data and instead look for patterns or data points that are significantly different from the rest of the data. These methods are useful when it is difficult or impossible to obtain labeled data, which is typically the case in computational workflows. Exampled methods like:
clustering (\eg, k-means \citep{lloyd1982least})
, density-based methods (\eg, k-nearest neighbors \citep{ramaswamy2000efficient})
, generative models (\eg, auto-encoder \citep{sakurada2014anomaly})
, etc.
In this work, we adopt the methods from PyOD \citep{zhao2019pyod}, where the data are considered as tabular data, and PyGOD \citep{liu2022bond}, where the data are considered as structural data, as benchmarks, reporting a set of metrics, and provide insightful analysis of the data we collected.

\vspace{-1em}
\paragraph{Benchmarking setup.}
We began by converting our parsed data from Section \ref{subsubsec:parsed_data} into a PyTorch Geometric (PyG) dataset \citep{Fey/Lenssen/2019}, where each graph $G$ is a DAG with nodes as jobs and edges as dependencies. PyG is widely used for graph data and can be converted to PyTorch and numpy tabular format for convenience in adapting existing models. We then normalized the datasets column-wise (feature) and aligned the timestamps. Finally, we randomly split the data into training, validation, and testing sets with a ratio of 7:1:2 with tabular, graph, and text data. A detailed data processing is provided in appendix \ref{appx:data_processing}.
The experiments were conducted on a single machine that featured two AMD EPYC 7532 processors, 256GB of local RAM, and an NVIDIA A100 GPU with a capacity of 40GB RAM. Because of the different backend implementations using either scikit-learn~\citep{scikit-learn} or PyTorch~\citep{paszke2019pytorch}, we give priority to using the GPU with the assistance of the CPU. More detailed hyperparameter setups of each algorithm are included in appendix \ref{appx:algorithm_setups}.

\begin{table}[t]
  \centering
  \caption{Unsupervised Model Examples}
  \label{tab:ul_results}
  \resizebox{\linewidth}{!}{
    \begin{tabular}{l||rrrr|r|rrrr|r}
      \toprule
      \multirow{2}{*}{Workflow} & \multicolumn{5}{c|}{GAE (PyGOD)}
                                & \multicolumn{5}{c}{GMM (PyOD)}
      \\
                                & Accuracy                         & AP   & ROC-AUC & Recall@k & Train time(sec.)
                                & Accuracy                         & AP   & ROC-AUC & Recall@k & Train time(sec.)
      \\
      \midrule
      1000 Genome               & .664                             & .334 & .523    & .131     & 4.950
                                & .660                             & .343 & .525    & .133     & 0.042
      \\
      CASA Nowcast              & .702                             & .213 & .467    & .048     & 39.390
                                & .767                             & .259 & .561    & .195     & 0.529
      \\
      CASA Wind Speed           & .709                             & .202 & .461    & .039     & 8.978
                                & .759                             & .227 & .538    & .160     & 0.085
      \\
      EHT Difmap                & .722                             & .214 & .491    & .085     & 2.204
                                & .748                             & .226 & .528    & .145     & 0.031
      \\
      EHT Imaging               & .730                             & .175 & .459    & .036     & 1.728
                                & .817                             & .267 & .603    & .265     & 0.020
      \\
      EHT Smili                 & .710                             & .195 & .446    & .010     & 1.847
                                & .761                             & .200 & .504    & .082     & 0.026
      \\
      Montage                   & .715                             & .199 & .466    & .045     & 12.533
                                & .753                             & .221 & .529    & .145     & 0.107
      \\
      Predict Future Sales      & .736                             & .183 & .478    & .063     & 6.725
                                & .773                             & .200 & .533    & .153     & 0.053
      \\
      PyCBC Inference           & .762                             & .181 & .508    & .114     & 1.924
                                & .781                             & .194 & .537    & .159     & 0.020
      \\
      PyCBC Search              & .722                             & .185 & .457    & .031     & 11.313
                                & .759                             & .196 & .516    & .126     & 0.108
      \\
      SOMOSPIE                  & .723                             & .191 & .468    & .047     & 3.168
                                & .761                             & .204 & .522    & .130     & 0.033
      \\
      Variant Calling           & .722                             & .210 & .490    & .085     & 8.832
                                & .763                             & .241 & .549    & .178     & 0.081
      \\
      \bottomrule
    \end{tabular}
  }
\end{table}

\begin{table}[t]
  \centering
  \caption{Supervised Model Examples}
  \label{tab:sl_results}
  \resizebox*{\linewidth}{!}{
    \begin{tabular}{l||rrrr|r|rrrr|r}
      \toprule
      \multirow{2}{*}{Workflow}
                           & \multicolumn{5}{c|}{GNN (GCN)}
                           & \multicolumn{5}{c}{LLM-SFT (BERT-base-uncased)}
      \\
                           & Accuracy                                        & AP   & ROC-AUC & Recall@k & Train time(sec.)
                           & Accuracy                                        & AP   & ROC-AUC & Recall@k & Train time(sec.)
      \\
      \midrule
      1000 Genome          & .794                                            & .376 & .555    & .150     & 98.153
                           & .776                                            & .537 & .746    & .664     & 158.524
      \\
      CASA Nowcast         & .784                                            & .217 & .502    & .007     & 2413.249
                           & .796                                            & .294 & .584    & .208     & 3215.485
      \\
      CASA Wind Speed      & .791                                            & .210 & .502    & .005     & 264.580
                           & .795                                            & .233 & .534    & .090     & 515.584
      \\
      EHT Difmap           & .786                                            & .215 & .502    & .006     & 27.223
                           & .784                                            & .282 & .583    & .231     & 54.185
      \\
      EHT Imaging          & .817                                            & .190 & .506    & .014     & 11.002
                           & .825                                            & .175 & .500    & .000     & 19.759
      \\
      EHT Smili            & .805                                            & .195 & .500    & .000     & 13.392
                           & .840                                            & .399 & .657    & .342     & 26.594
      \\
      Montage              & .795                                            & .207 & .502    & .006     & 1092.069
                           & .803                                            & .240 & .538    & .091     & 572.677
      \\
      Predict Future Sales & .814                                            & .186 & .500    & .000     & 184.743
                           & .861                                            & .394 & .667    & .359     & 290.684
      \\
      PyCBC Inference      & .815                                            & .185 & .500    & .000     & 15.316
                           & .829                                            & .222 & .531    & .066     & 27.950
      \\
      PyCBC Search         & .809                                            & .191 & .504    & .014     & 421.313
                           & .825                                            & .255 & .571    & .166     & 610.556
      \\
      SOMOSPIE             & .801                                            & .199 & .500    & .000     & 58.074
                           & .809                                            & .279 & .606    & .276     & 98.941
      \\
      Variant Calling      & .788                                            & .214 & .505    & .014     & 275.890
                           & .791                                            & .252 & .552    & .139     & 483.320
      \\
      \bottomrule
    \end{tabular}
  }
\end{table}

\vspace{-1em}
\paragraph{Metrics and benchmark results.}
Table~\ref{tab:dataset} shows that the binary labels in the dataset are imbalanced. Therefore, performance is evaluated using metrics such as ROC-AUC, average precision, and top-k recall scores, in which we initially set the value of $k$ to the number of true anomalies in the test set. In addition, we also provide the accuracy score for both supervised and unsupervised approaches for comparison across different methods.

The experimental evaluation encompassed the twelve novel scientific workflow applications introduced in this work, employing both supervised (Table~\ref{tab:sl_results}) and unsupervised (Table~\ref{tab:ul_results}) learning methodologies. The reported results constitute the mean values across five independent runs. Due to space constraints, comprehensive descriptions of the benchmarks, hyperparameter configurations, and complete results are provided in Appendix \ref{appx:benchmark_results}. In Table~\ref{tab:ul_results}, we present a comparative analysis of the Graph Auto-encoder (GAE) and Gaussian Mixture Model (GMM) architectures, which operate on graph and tabular data formats, respectively. It is noteworthy that the GAE approach exhibited significantly longer training times, despite leveraging GPU acceleration, without yielding substantial performance improvements over the established methods implemented in the PyOD library. The inefficiency observed in handling large-scale workflows can be attributed to the computational complexity associated with operations such as graph completion and reconstruction, which operate on sparse graph representations.

In Table~\ref{tab:sl_results}, we present a comparative evaluation of supervised learning approaches, including Graph Convolutional Networks (GCN) and Supervised Fine-Tuned (SFT) models based on the BERT architecture with Low-Rank Adaptation (LoRA) \citep{hu2021lora} for parameter-efficient training, resulting in training with 0.72\% of total parameters in BERT-base-uncased pre-trained model. The GCN models operate on graph-structured data, while the SFT models leverage textual inputs. It is observed that both methodologies necessitate substantial training efforts due to the inherent complexity of the underlying architectures. Notably, the SFT models based on pre-trained model, despite their relatively compact size, achieves superior performance on certain workflows (e.g., EHT Smili) compared to their GCN counterparts. The textual data, coupled with the knowledge distilled from large-scale pre-training corpora, enables the LLMs to capture not only the topological dependencies encoded in the graph representations but also the intricate system configurations present in the textual descriptions. This synergistic approach facilitates a more comprehensive understanding of the underlying job requirements within the scientific workflows.

Given the results from benchmarking methods, there is a clear need to develop novel methods to advance the state of the art and improve the performance of anomaly detection methods that scales to large-scale scientific workflows. The results also highlight the importance of developing methods that can handle the complexity and heterogeneity of computational workflows, as well as the need for more research in this area to improve the reliability and reproducibility of computational workflows.

\vspace{-1em}
\section{Generalization and Limitations}
\vspace{-1em}
\label{sec:generalization_limitation}

In this work, we provide twelve workflows, eleven of which correspond to scientific applications and one of which corresponds to a data science workflow.
While it is important to develop novel methods to facilitate unsupervised anomaly detection on these workflows, the use of these workflows in developing methods for a larger set of workflow anomaly detection problems is also relevant. Therefore, an important aspect of this work is the ability of the benchmark dataset to simulate anomalies across different scientific domains.

Since the workflow data is modeled as a DAG, it is feasible to use the ML methods that are developed to be precise on these workflows to work with other datasets that can be represented as DAG. Therefore, one may design approaches that can be built for anomaly detection for any application that generates data represented as a DAG, including workflows that are not represented in this benchmark since a wide range of anomalies can occur in real-world computational workflows.

Moreover, as we provide the raw log data, researchers can use it to explore new methods and approaches for anomaly detection beyond the ones presented in this work. This can lead to the development of more accurate and robust anomaly detection algorithms that can be applied to a wider range of computational workflows. For example, taking the context of the raw logs as inputs for natural language processing~(NLP) tasks, such as large language models, can enable the development of more advanced anomaly detection algorithms that can capture the complex relationships between the various components of computational workflows, beyond the scope of graphs only capture the dependence between jobs.

Despite their usefulness, benchmark datasets with synthetic anomalies in computational workflows have some limitations. One limitation is that they may not fully capture the complexity and variability of real-world computational workflows. Scientific and data science workflows can be highly heterogeneous and may contain anomalies that are difficult to simulate using synthetic data. Another limitation is that benchmark datasets may not reflect the specific characteristics of the data in a particular scientific domain. Anomaly detection algorithms that perform well on benchmark datasets may not necessarily perform well on real-world data in a different scientific domain. Therefore, it is important to evaluate the performance of anomaly detection algorithms on real-world computational workflows in addition to benchmark datasets.

\vspace{-1em}
\section{Conclusion}
\vspace{-1em}
In this paper, we developed benchmark datasets for anomaly detection in computational workflows. Specifically, we have introduced a new dataset for benchmarking anomaly detection techniques, which includes systematically injected anomalies and raw execution logs from workflows executing on distributed infrastructures. We have also summarized the statistics of new datasets, including twelve computational workflows.

Our benchmarking, including both supervised and unsupervised approaches, provides effective methods for detecting anomalies in computational workflows. However, we also highlight the need for more research in this area, particularly in developing techniques that can handle the complexity and heterogeneity of computational workflows. We hope the study provides valuable insights into the challenges of anomaly detection in computational workflows and offers a starting point for future research in this area. We hope that our benchmark dataset and analysis will be useful for researchers and practitioners working on improving the reliability and reproducibility of computational workflows.

\newpage
\begin{ack}
  \vspace{-1em}
  This work is funded by the Department of Energy under the Integrated Computational and Data Infrastructure (ICDI) for Scientific Discovery, grant DE-SC0022328.
\end{ack}

{
\bibliography{ref.bib}
\bibliographystyle{plainnat}
}

\medskip
\newpage


\appendix
\section*{Appendix}

FlowBench code and data repository: \url{https://github.com/PoSeiDon-Workflows/flowbench}.
We also provide our documentation: \url{https://poseidon-workflows.github.io/FlowBench/}.

\section{Workflow Execution Infrastructure}
\label{appx:workflow_execution_infra}
\begin{figure}[h]
  \centering
  \includegraphics[width=\linewidth]{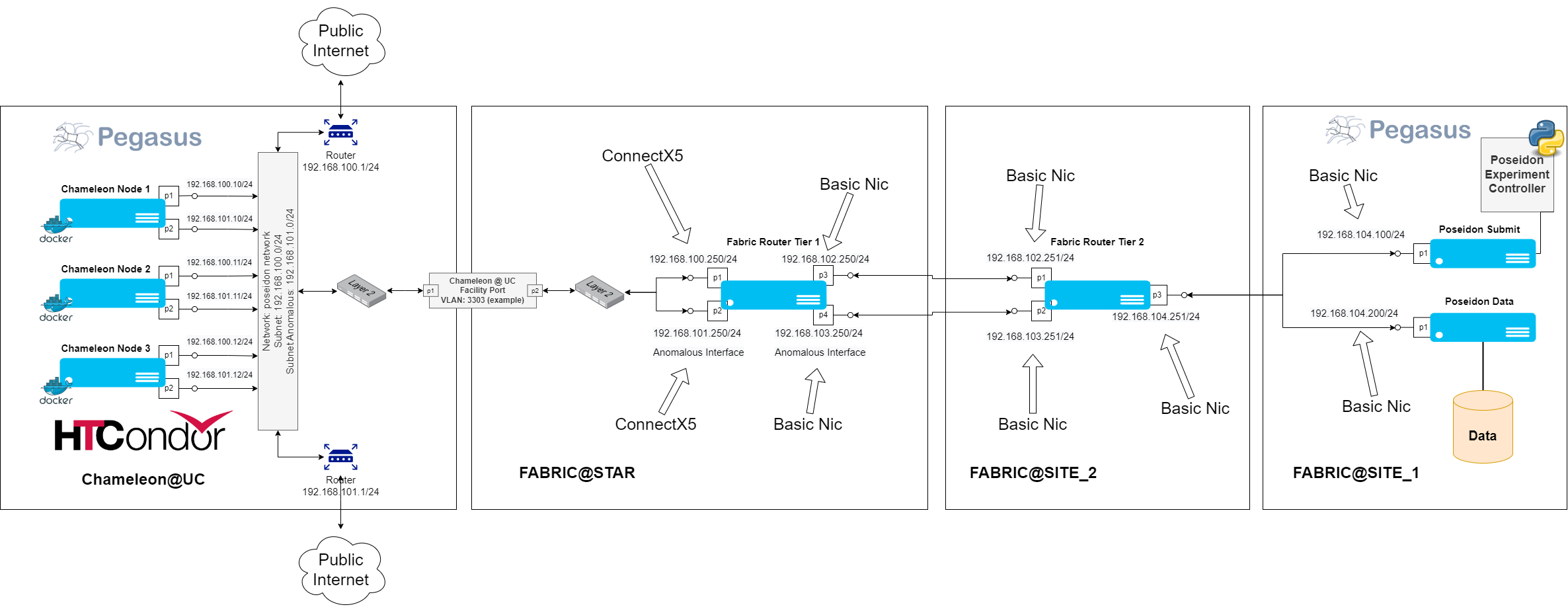}
  \caption{\footnotesize Overview of the execution infrastructure. The deployment spans across the Chameleon Cloud and the FABRIC testbed. Chameleon hosts the workers while FABRIC hosts the networking infrastructure, the workflow submission node and the data storage node. Docker containers are deployed on baremetal nodes and interference (CPU, HDD) is introduced using cgroups. An experimental controller on the workflow submission node orchestrates the anomaly injection, workflow execution triggering and data labeling.}
  \label{fig:poseidon-deployment}
\end{figure}

\section{Workflow Diagrams}
\label{appx:workflow_diagrams}

\begin{figure}[h]
  \centering
  \includegraphics[width=\linewidth]{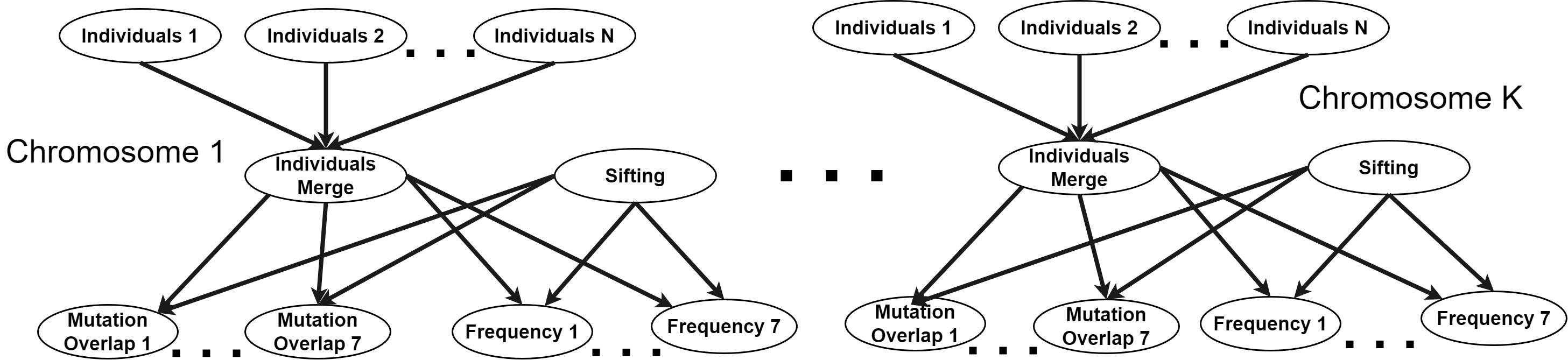}
  \caption{\footnotesize Overview of the 1000Genome sequencing analysis workflow. The workflow creates a branch for each chromosome and each individual task is processing a subset of the Phase 3 data (equally distributed).}
  \label{fig:1000genome}
\end{figure}

\begin{figure}[h]
  \centering
  \includegraphics[width=\linewidth]{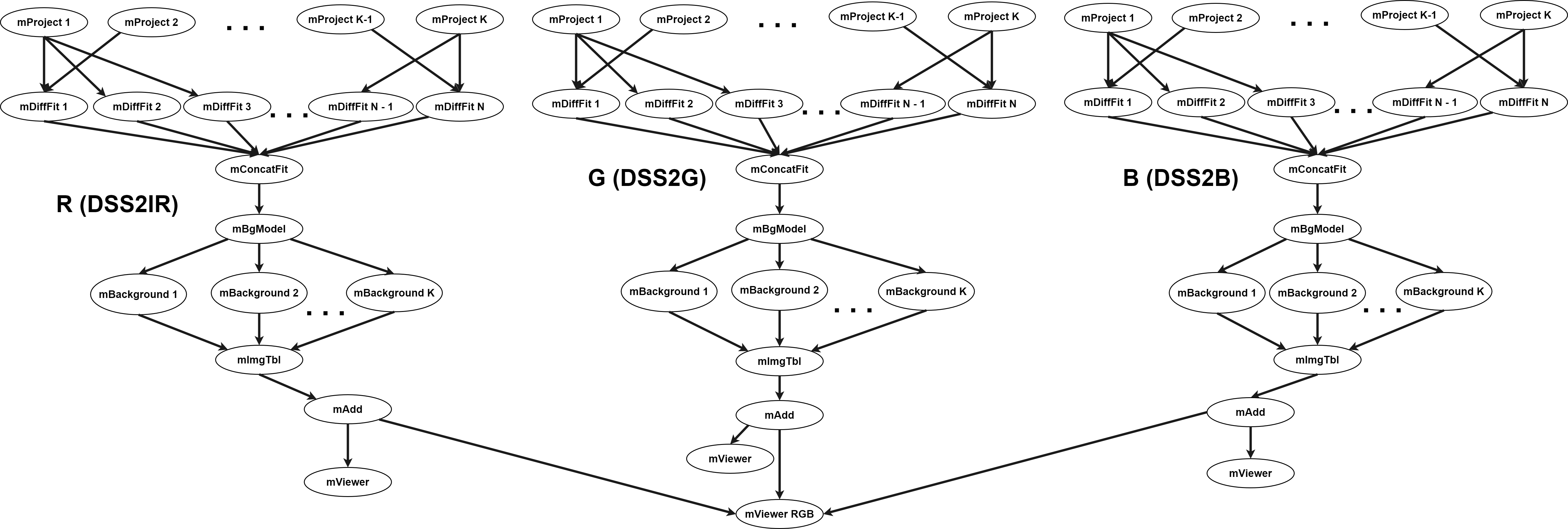}
  \caption{\footnotesize Overview of the Montage workflow. In this case, the workflow uses images captured by the Digitized Sky Survey (DSS)~\citep{dss-archive} and creates a branch for each band that is requested to be processed during the workflow generation. The size of the first level of each branch depends on the size of the section of the sky to be analyzed, while the second level on the number of overlapping images stored in the archive.}
  \label{fig:montage}
\end{figure}

\begin{figure}[h]
  \centering
  \includegraphics[width=\linewidth]{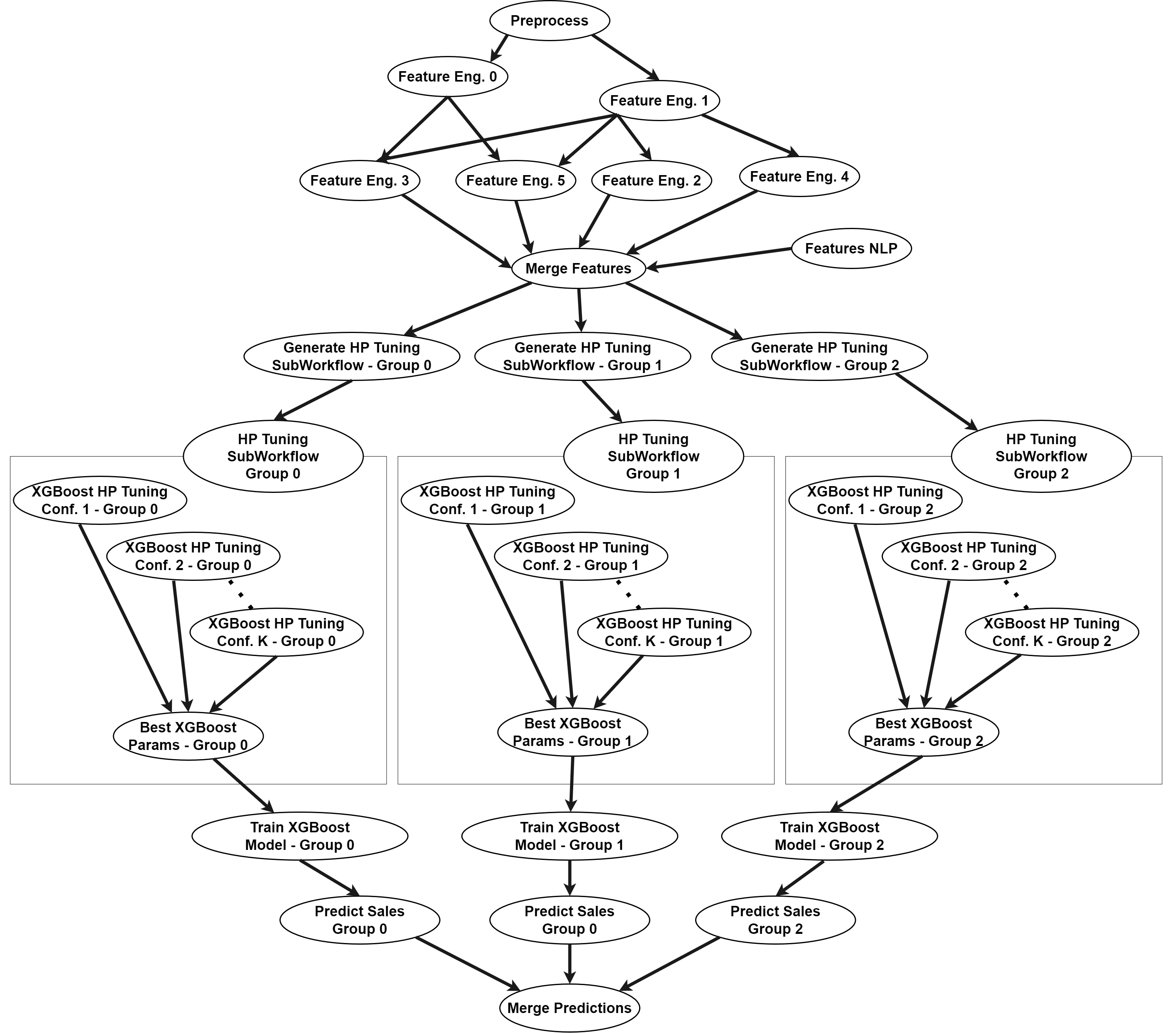}
  \caption{\footnotesize Overview of the Predict Future Sales workflow. The workflow splits the data into 3 item categories and trains 3 XGBoost models that are later combined, using an ensemble technique. It contains 3 hyperparameter tuning subworkfows, that test different sets of features and picks the best performing one. The number of HPO tasks is configurable and depends on the number of combinations that will be tested.}
  \label{fig:predict-future-sales}
\end{figure}

Figure~\ref{fig:1000genome}, \ref{fig:montage} and \ref{fig:predict-future-sales} are the workflow diagrams of the 1000 Genome, Montage and Predict Future Sales workflows, respectively.

\begin{figure}[h]
  \centering
  \includegraphics[width=\linewidth]{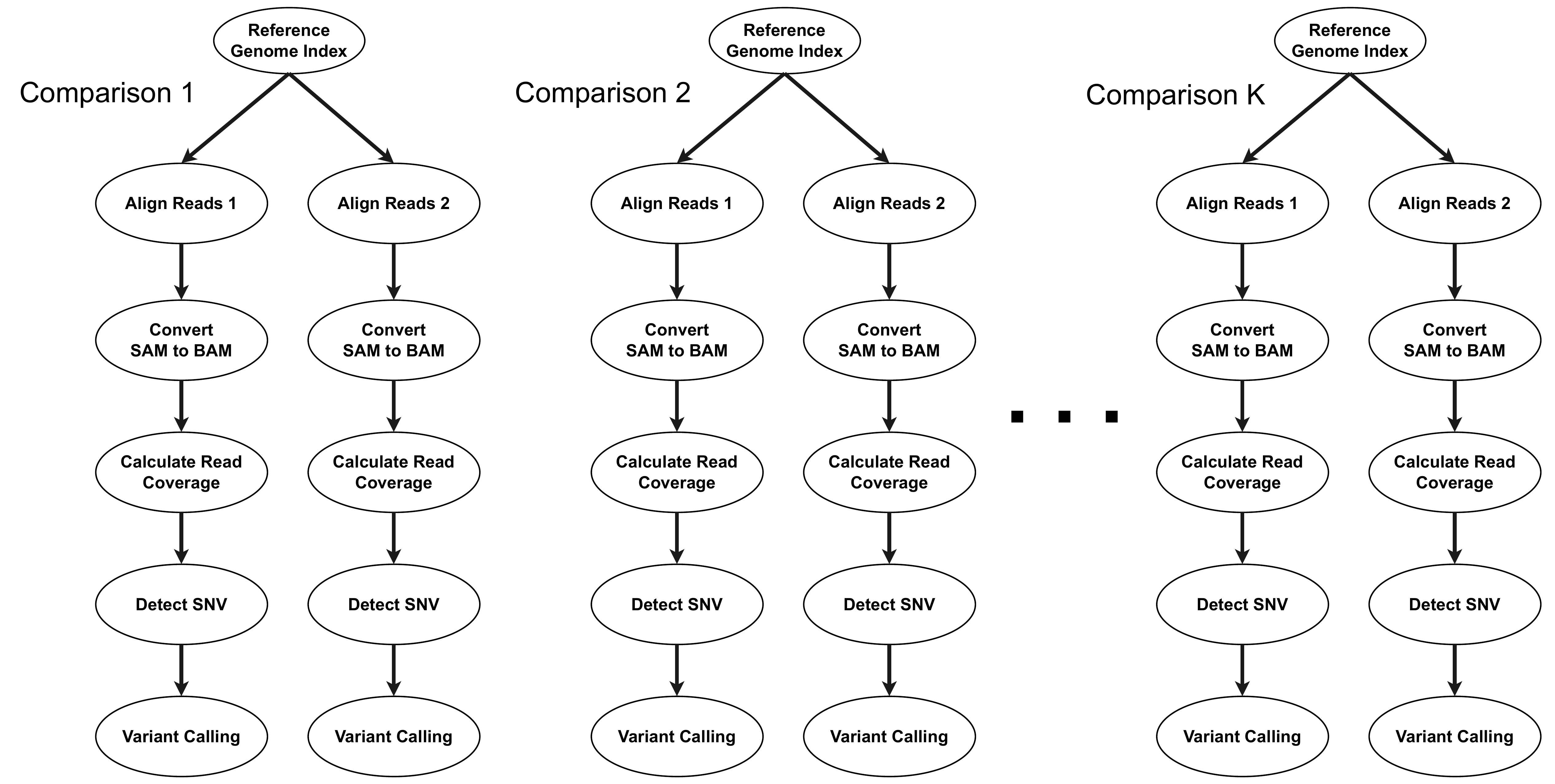}
  \caption{\footnotesize Overview of the Variant Calling Workflow. The workflow compares a DNA sequence to a reference genome and performs multiple comparisons in parallel.}
  \label{fig:variant-calling}
\end{figure}

\begin{figure}[h]
  \centering
  \includegraphics[width=\linewidth]{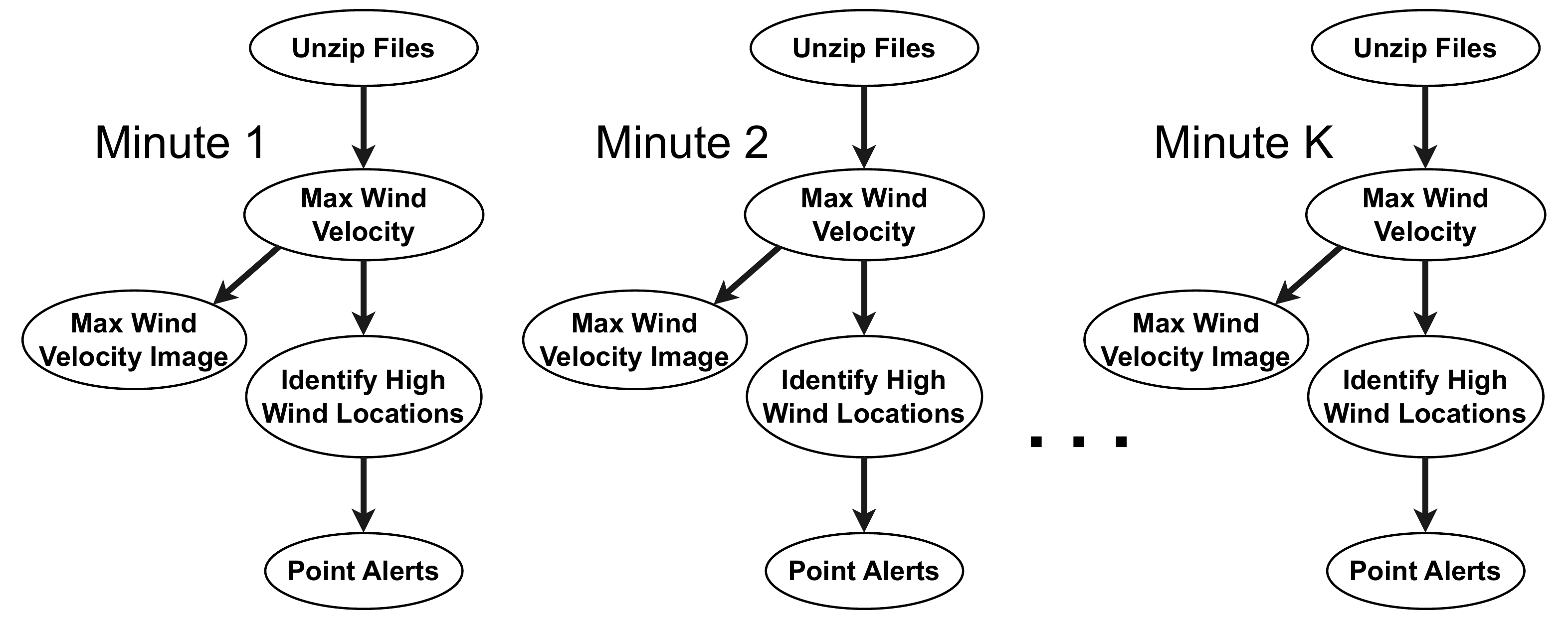}
  \caption{\footnotesize Overview of the CASA Wind Speed Workflow. The wind speed workflows calculates the max wind velocity in increments of one minute. Multiple minutes are calculated in parallel.}
  \label{fig:casa-wind}
\end{figure}

\begin{figure}[h]
  \centering
  \includegraphics[width=\linewidth]{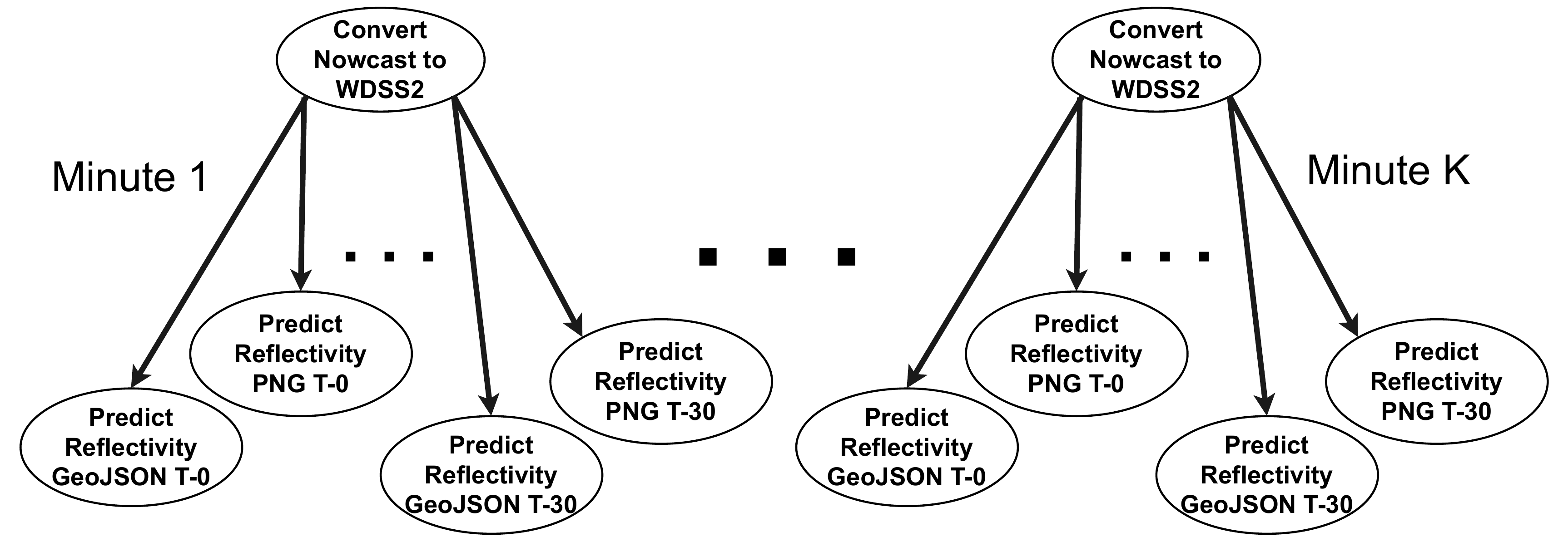}
  \caption{\footnotesize Overview of the CASA Nowcast Workflow. The nowcast workflow forecasts the nowcast events for the next thirty minutes. The forecats are based on one minute radar data. Multiple minutes of data are calculated in parallel.}
  \label{fig:casa-nowcast}
\end{figure}

\begin{figure}[h]
  \centering
  \includegraphics[width=\linewidth]{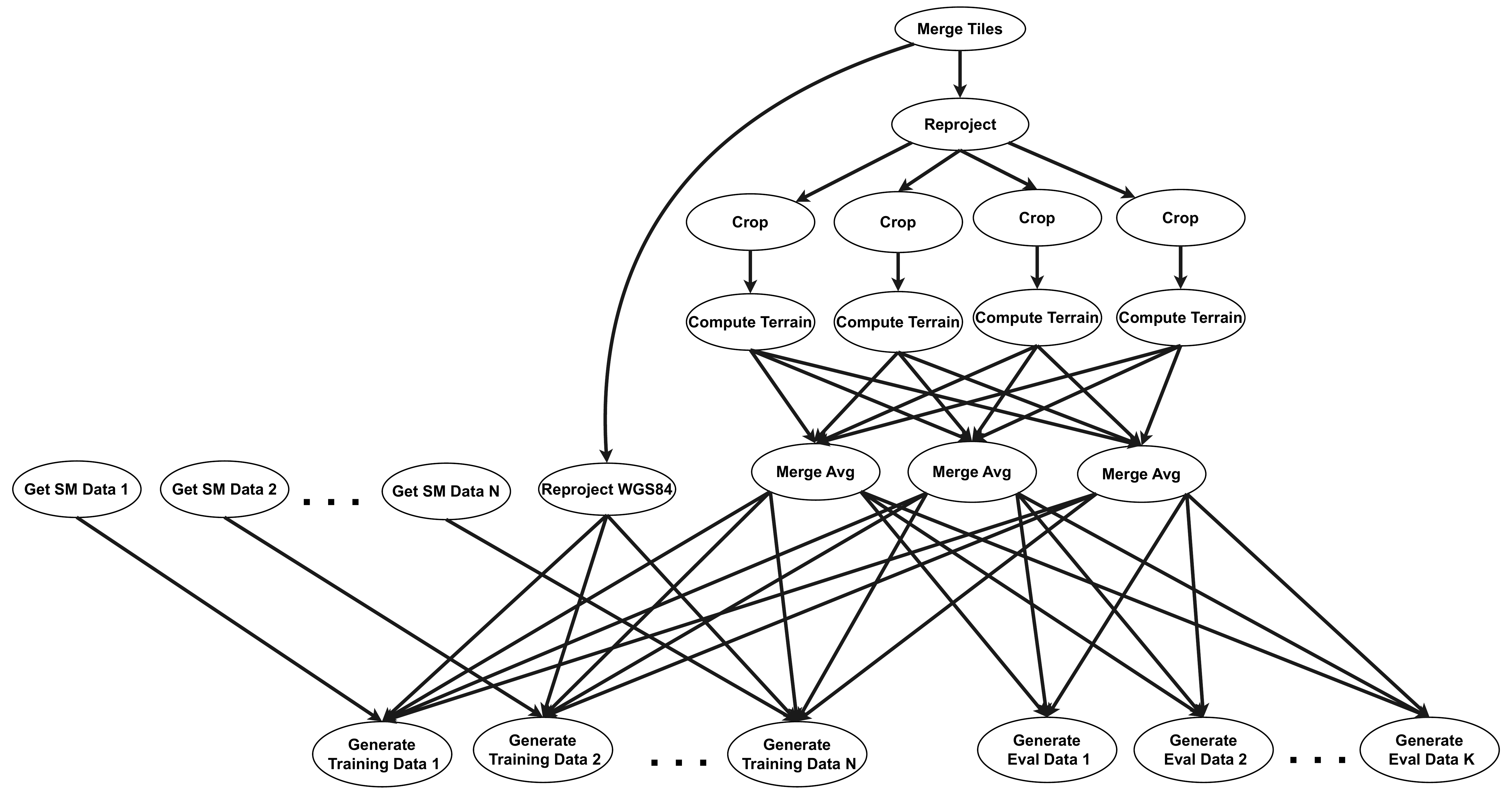}
  \caption{\footnotesize Overview of the Soil Moisture Spatial Inference Engine Workflow - Data Preparation. The workflow fetches and process environmental data for a specified region using USGS data.}
  \label{fig:somospie}
\end{figure}

\begin{figure}[h]
  \centering
  \includegraphics[width=\linewidth]{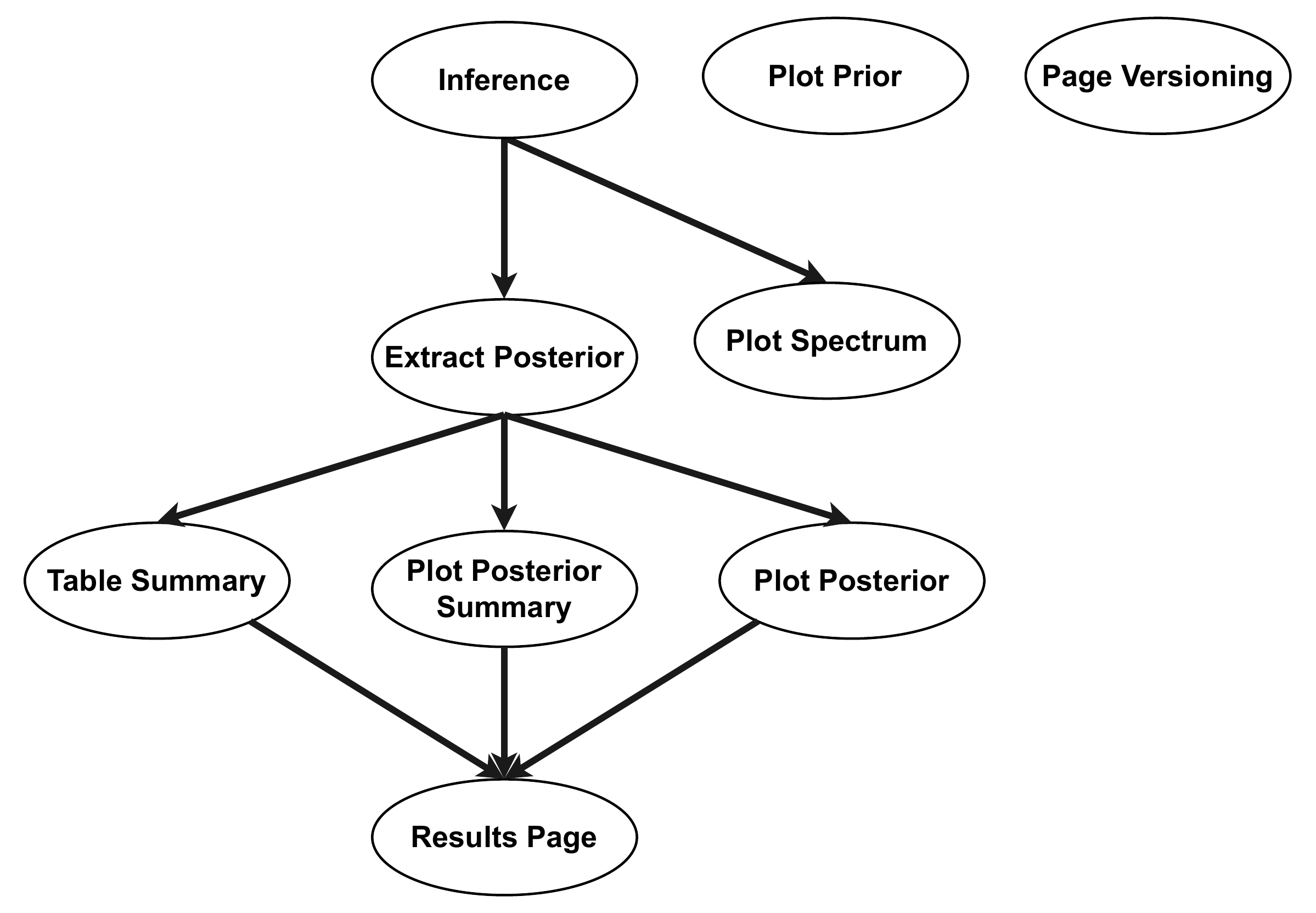}
  \caption{\footnotesize Overview of the PyCBC Inference Workflow. The inference workflow performs parameter estimation on gravitational wave signals using prior knowledge and Bayesian approaches.}
  \label{fig:pycbc-inference}
\end{figure}

\begin{figure}[h]
  \centering
  \includegraphics[width=\linewidth]{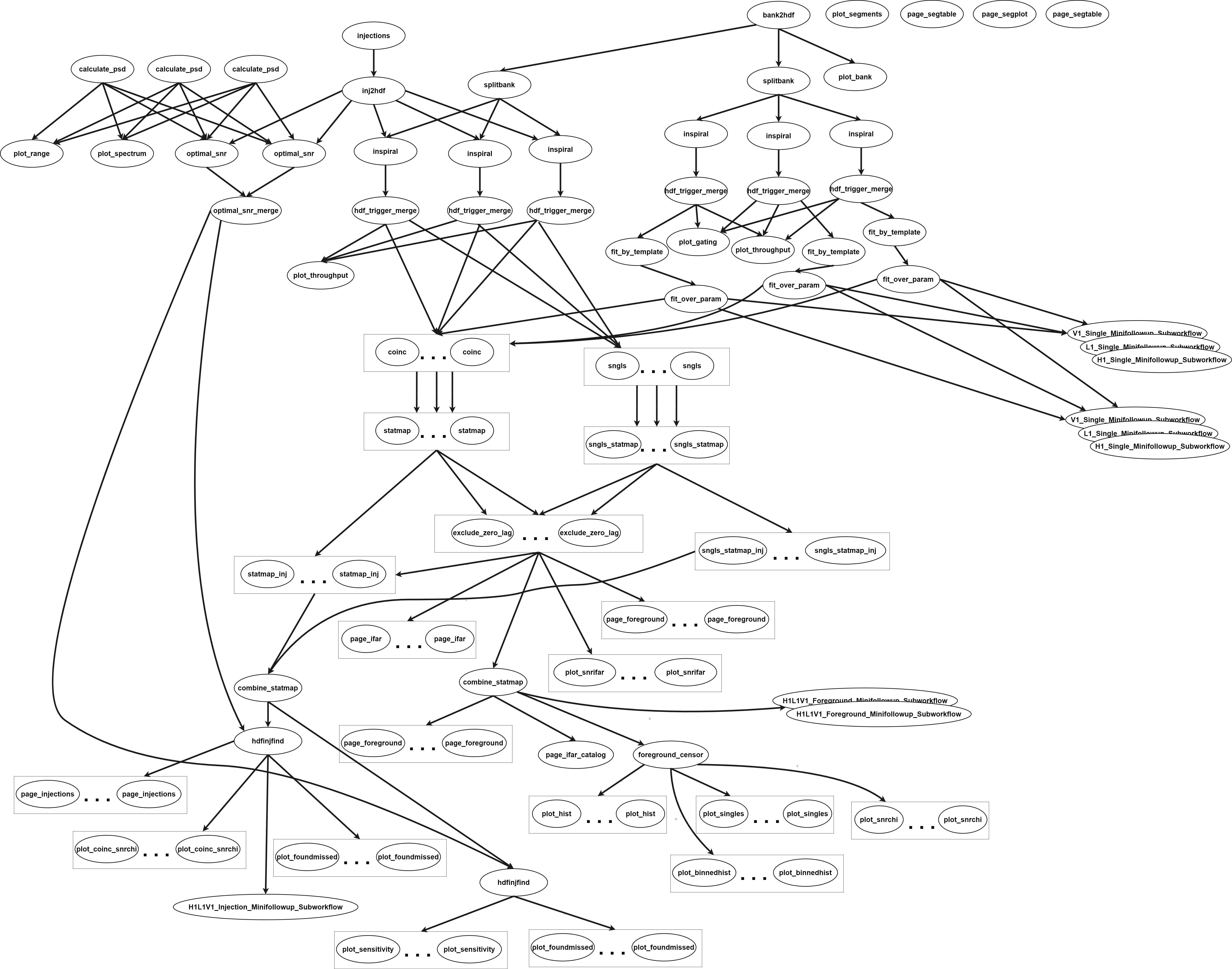}
  \caption{\footnotesize Overview of the PyCBC Search Workflow. The search workflow sifts through data to identify gravitational waves. It contains nine subworkflows for followup studies.}
  \label{fig:pycbc-search}
\end{figure}

\begin{figure}[h]
  \centering
  \includegraphics[width=\linewidth]{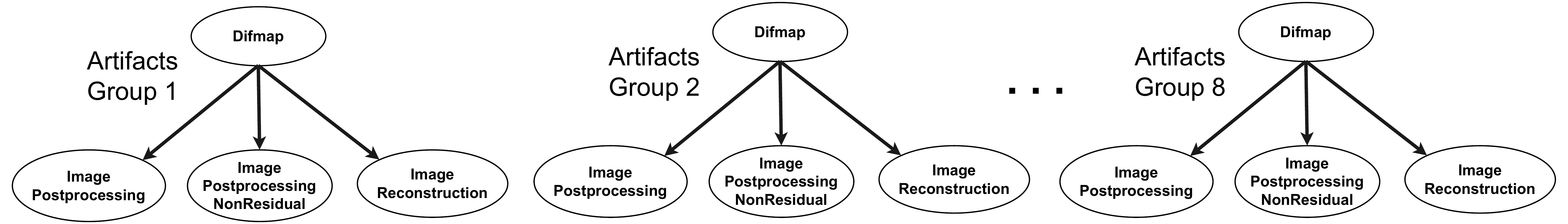}
  \caption{\footnotesize Overview of the EHT Difmap Workflow. The workflow performs image reconstruction by employing iterative convolution with difference mapping.}
  \label{fig:eht-difmap}
\end{figure}

\begin{figure}[h]
  \centering
  \includegraphics[width=\linewidth]{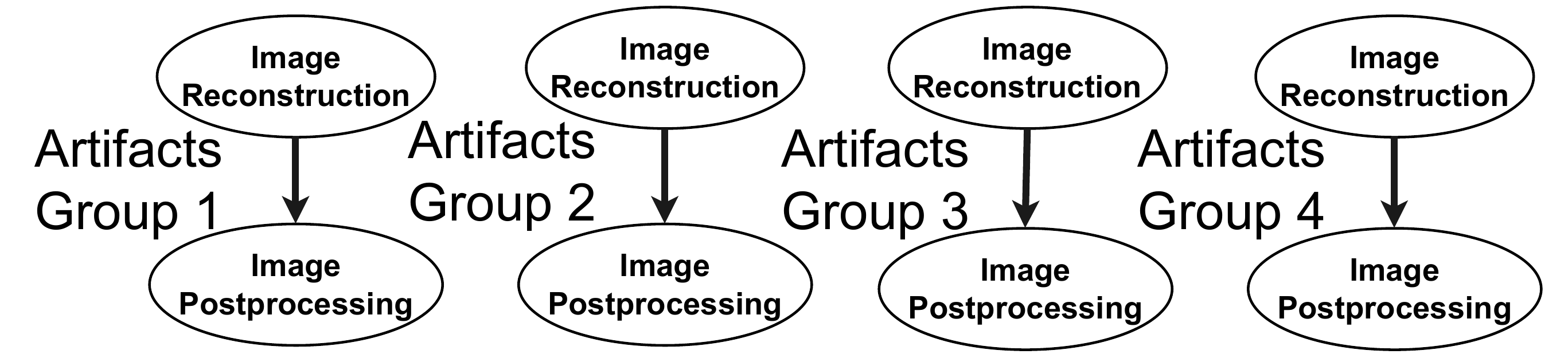}
  \caption{\footnotesize Overview of the EHT Imaging Workflow. The workflow performs image reconstruction using the regularized maximum likelihood (RLM) approach.}
  \label{fig:eht-imaging}
\end{figure}

\begin{figure}[h]
  \centering
  \includegraphics[width=\linewidth]{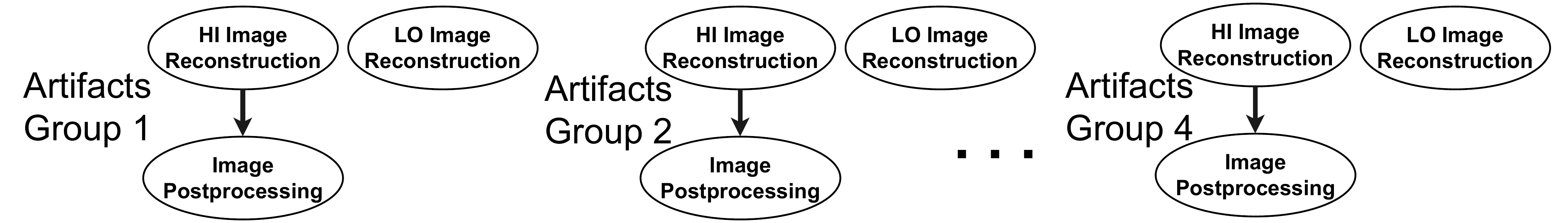}
  \caption{\footnotesize Overview of the EHT Smili Workflow. The workflow uses the Sparse Modeling Imaging Library for Interferometry (SMILI) to incorporate pre-calibrated datasets and perform image reconstruction using the RML approach.}
  \label{fig:eht-smili}
\end{figure}

\section{Parsed Features}
\label{appx:parsed_features}

\begin{figure}[h]
  \centering
  \includegraphics[width=0.8\linewidth]{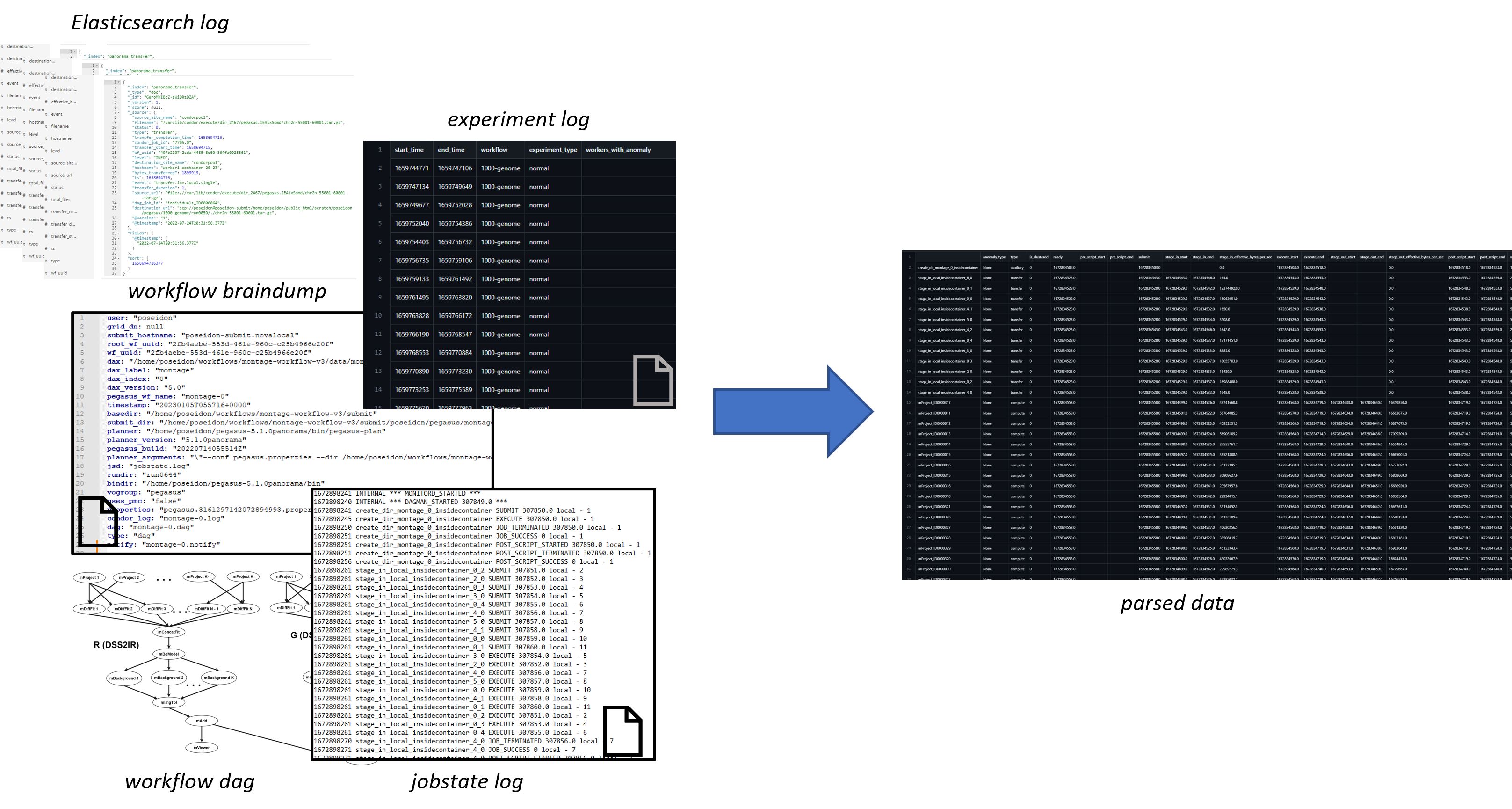}
  \caption{\footnotesize Converting the raw data to the parsed CSV data}
  \label{fig:raw-data-parsing}
\end{figure}

Table~\ref{tab:parsed-features} shows the features that are parsed from the Pegasus workflow logs, with the type of each feature and a brief description.
\begin{table}[h]
  \centering
  \caption{Parsed Dataset Features.}
  {
    \label{tab:parsed-features}
    \resizebox{\linewidth}{!}{
      \begin{tabular}{l|l|l}
        \toprule
        Field                                          & Data Type & Description                                                                                       \\
        \midrule
        node\_id                                       & string    & Exec. DAG Node ID                                                                                 \\
        anomaly\_type                                  & string    & Anomaly label                                                                                     \\
        type                                           & int       & Pegasus job type                                                                                  \\
        is\_clustered                                  & int       & 1 if job clustering enabled else 0                                                                \\
        ready                                          & ts        & Epoch ts when all dependencies have been met and job can be dispatched                            \\
        pre\_script\_start                             & ts        & Epoch ts when the pre-script started executing                                                    \\
        pre\_script\_end                               & ts        & Epoch ts when the pre-script stopped executing                                                    \\
        submit                                         & ts        & Epoch ts when the job was submitted to the queue                                                  \\
        stage\_in\_start                               & ts        & Epoch ts when the data stage in started                                                           \\
        stage\_in\_end                                 & ts        & Epoch ts when the data stage in ended                                                             \\
        stage\_in\_effective\_bytes\_per\_sec          & float     & Bytes written per second (input data)                                                             \\
        execute\_start                                 & ts        & Epoch ts when the execution starts                                                                \\
        execute\_end                                   & ts        & Epoch ts when the execution ends                                                                  \\
        stage\_out\_start                              & ts        & Epoch ts when the data stage out started                                                          \\
        stage\_out\_end                                & ts        & Epoch ts when the data stage out ended                                                            \\
        stage\_out\_effective\_bytes\_per\_sec         & float     & Bytes written per second (output data)                                                            \\
        post\_script\_start                            & ts        & Epoch ts when the post-script started executing                                                   \\
        post\_script\_end                              & ts        & Epoch ts when the post-script ended executing                                                     \\
        wms\_delay                                     & float     & Composite field estimating the delay introduced by the WMS while preparing the job for submission \\
        pre\_script\_delay                             & float     & Composite field estimating the delay introduced by the pre-script                                 \\
        queue\_delay                                   & float     & Composite field estimating the time spent in the queue                                            \\
        runtime                                        & float     & Total runtime of the job, based on execute start and end                                          \\
        post\_script\_delay                            & float     & Composite field estimating the delay introduced by the post-script                                \\
        stage\_in\_delay                               & float     & Total time spend staging in data, based on stage in start and end                                 \\
        stage\_in\_bytes                               & float     & Total bytes staged in                                                                             \\
        stage\_out\_delay                              & float     & Total time spend staging out data, based on stage out start and end                               \\
        stage\_out\_bytes                              & float     & Total bytes staged out                                                                            \\
        kickstart\_user                                & string    & Name of the user submitted the job                                                                \\
        kickstart\_site                                & string    & Name of the execution site                                                                        \\
        kickstart\_hostname                            & string    & Hostname of the worker node the job executed on                                                   \\
        kickstart\_transformations                     & string    & Mapping of the executable locations                                                               \\
        kickstart\_executables                         & string    & Names of the invoked executables                                                                  \\
        kickstart\_executables\_argv                   & string    & Command line arguments used to invoke the executables                                             \\
        kickstart\_executables\_cpu\_time              & float     & Total cpu time                                                                                    \\
        kickstart\_status                              & int       & The status of the job as marked by Pegasus Kickstart                                              \\
        kickstart\_executables\_exitcode               & int       & The exitcode of the invoked executable(s)                                                         \\
        kickstart\_online\_iowait                      & float     & Time spent on waiting for io (seconds)                                                            \\
        kickstart\_online\_bytes\_read                 & float     & Total bytes read from disk (bytes)                                                                \\
        kickstart\_online\_bytes\_written              & float     & Total bytes written to disk (bytes)                                                               \\
        kickstart\_online\_read\_system\_calls         & float     & Number of read system calls                                                                       \\
        kickstart\_online\_write\_system\_calls        & float     & Number of write system calls                                                                      \\
        kickstart\_online\_utime                       & float     & Time spent on user space                                                                          \\
        kickstart\_online\_stime                       & float     & Time spent on kernel space                                                                        \\
        kickstart\_online\_bytes\_read\_per\_second    & float     & Bytes read per second (effective)                                                                 \\
        kickstart\_online\_bytes\_written\_per\_second & float     & Bytes written per second (effective)                                                              \\
        \bottomrule
      \end{tabular}
    }}
\end{table}

\section{Benchmark algorithms}
\label{appx:algorithms}
Table~\ref{tab:algorithms} provide the selected benchmark algorithms and their references.
The input $X$ represents using features only, while $A, X$ represents using both structure and features.
\begin{table}[t]
  \centering
  \caption{Selected algorithms for anomaly detection.
    The input $X$ represents using features only, while $A, X$ represents using both structure and features.}
  \label{tab:algorithms}
  \resizebox{0.85\columnwidth}{!}{
    \begin{tabular}{rl|ccc}
      \toprule
      Algorithm     & Reference                       & Input Data & ML Supervision  & Backbone          \\
      \midrule
      DT            & \cite{quinlan1986induction}     & X          & Supervised      & Decision Tree     \\
      RF            & \cite{breiman2001random}        & X          & Supervised      & Random Forest     \\
      MLP           & \cite{rumelhart1986learning}    & X          & Supervised      & MLP               \\
      GCN           & \cite{kipf2017semi}             & A, X       & Supervised      & GNN               \\
      GraphSAGE     & \cite{hamilton2017inductive}    & A, X       & Supervised      & GNN               \\
      FTTransformer & \cite{gorishniy2021revisiting}  & X          & Supervised      & Transformer       \\
      LLMs (SFT)    & \cite{DBLP:abs-1810-04805}      & X (text)   & Supervised      & Transformer       \\
      \midrule
      GANomaly      & \cite{akcay2019ganomaly}        & X          & Semi-supervised & GAN               \\
      DeepSAD       & \cite{ruff2019deep}             & X          & Semi-supervised & Deep one-class    \\
      REPEN         & \cite{pang2019deep}             & X          & Semi-supervised & Neighbors         \\
      PReNet        & \cite{ren2019progressive}       & X          & Semi-supervised & Neighbors         \\
      \midrule
      ABOD          & \cite{kriegel2008angle}         & X          & Unsupervised    & Probabilistic     \\
      CBLOF         & \cite{he2003discovering}        & X          & Unsupervised    & Neighbors         \\
      FB            & \cite{lazarevic2005feature}     & X          & Unsupervised    & Ensembles         \\
      HBOS          & \cite{goldstein2012histogram}   & X          & Unsupervised    & Neighbors         \\
      IF            & \cite{liu2008isolation}         & X          & Unsupervised    & Ensembles         \\
      KNN           & \cite{ramaswamy2000efficient}   & X          & Unsupervised    & Neighbors         \\
      AKNN          & \cite{angiulli2002fast}         & X          & Unsupervised    & Neighbors         \\
      LOF           & \cite{breunig2000lof}           & X          & Unsupervised    & Neighbors         \\
      MCD           & \cite{hardin2004outlier}        & X          & Unsupervised    & Linear            \\
      OCSVM         & \cite{scholkopf2001estimating}  & X          & Unsupervised    & Linear            \\
      PCA           & \cite{shyu2003novel}            & X          & Unsupervised    & Decomposition     \\
      LSCP          & \cite{zhao2019lscp}             & X          & Unsupervised    & Ensembles         \\
      INNE          & \cite{bandaragoda2018isolation} & X          & Unsupervised    & Ensembles         \\
      GMM           & \cite{rasmussen1999infinite}    & X          & Unsupervised    & Mixture           \\
      KDE           & \cite{latecki2007outlier}       & X          & Unsupervised    & Probabilistic     \\
      LMDD          & \cite{arning1996linear}         & X          & Unsupervised    & Linear            \\
      MLPAE         & \cite{sakurada2014anomaly}      & X          & Unsupervised    & MLP, AE           \\
      SCAN          & \cite{xu2007scan}               & A, X       & Unsupervised    & Clustering        \\
      Radar         & \cite{li2017radar}              & A, X       & Unsupervised    & Decomposition     \\
      Anomalous     & \cite{peng2018anomalous}        & A, X       & Unsupervised    & Decomposition     \\
      GCNAE         & \cite{kipf2016variational}      & A, X       & Unsupervised    & GNN, Auto-encoder \\
      Dominant      & \cite{ding2019deep}             & A, X       & Unsupervised    & GNN, Auto-encoder \\
      DONE          & \cite{bandyopadhyay2020outlier} & A, X       & Unsupervised    & MLP, Auto-encoder \\
      ADONE         & \cite{bandyopadhyay2020outlier} & A, X       & Unsupervised    & MLP, Auto-encoder \\
      AnomalyDAE    & \cite{fan2020anomalydae}        & A, X       & Unsupervised    & GNN, Auto-encoder \\
      GAAN          & \cite{chen2020generative}       & A, X       & Unsupervised    & GAN               \\
      GUIDE         & \cite{yuan2021higher}           & A, X       & Unsupervised    & GNN, Auto-encoder \\
      CONAD         & \cite{xu2022contrastive}        & A, X       & Unsupervised    & GNN, Auto-encoder \\
      \bottomrule
    \end{tabular}
  }
\end{table}

\section{Data processing}
\label{appx:data_processing}
Followed by Table~\ref{tab:parsed-features}, we further processed the parsed data into a dataset in PyTorch-Geometric (PyG) format~\citep{Fey/Lenssen/2019} for the graph-based algorithms.
The PyG format is a standard format for graph data in deep learning, which is widely used in the graph neural network (GNN) community.
The PyG format is a tuple of $(\mathbf{X}, \mathbf{A}, \mathbf{y})$, where $\mathbf{X}$ is the feature matrix, $\mathbf{A}$ is the adjacency matrix, and $\mathbf{y}$ is the label vector.
First, for the features with integers/strings as categories, we processed them into one-hot encoding, and then concatenate them into a feature vector $\Xvec$.
Notice that there are multiple features with timestamps, we shifted the timestamp to the same starting point (i.e., the first timestamp in the dataset), which allows us to measure the behavior of each individual job independently.
Moreover, we normalized job feature column-wise, in which we use min-max normalization such that normalized features are in the range $\sbr{0, 1}$.
Second, for the structural information, we convert the dependencies as directed edges between nodes, and furthermore, we convert the directed acyclic graphs into undirected graphs as $\Avec$.
Finally, we use the label vector $\yvec$ to indicate job anomalies, where $1$ indicates anomalous jobs and $0$ indicates normal jobs.
For the experiment in the paper, we split the dataset into training, validation, and test set with a ratio $0.6$, $0.2$, and $0.2$, and use the dataloader to load the dataset in mini-batch for training. All the steps are built into a pipeline as the transform class in PyG, which allows us to easily process the dataset into PyG format.

\section{Detailed benchmark settings for algorithms}
\label{appx:algorithm_setups}
Following in Table~\ref{tab:algorithms}, we provide the detailed settings for each algorithm used in the benchmark. Notice that we use the default settings for all the algorithms, and we do not tune the hyperparameters for each algorithm. For the hyperparameters that are not listed below, we use the default settings in the implementation of each algorithm from PyOD and PyGOD.

\begin{itemize}
  \item {\bf ABOD}. N/A.
  \item {\bf CBLOF}. Check estimator: False.
  \item {\bf FB}. Number of neighbors: 35.
  \item {\bf HBOS}. N/A.
  \item {\bf IF}. N/A.
  \item {\bf KNN}. N/A.
  \item {\bf AKNN}. Aggregation method: average.
  \item {\bf LOF}. Number of neighbors: 35.
  \item {\bf MCD}. N/A.
  \item {\bf OCSVM}. N/A.
  \item {\bf PCA}. N/A.
  \item {\bf LSCP}. Detector list with a set of LOF models on a number of neighbors: 5, 10, 15, 20, 25, 30, 35, 40, 45, 50.
  \item {\bf INNE}. Max samples: 2.
  \item {\bf GMM}. N/A.
  \item {\bf KDE}. N/A.
  \item {\bf LMDD}. N/A.
  \item {\bf MLPAE}. Hidden dimension: 64, batch size: 64 learning rate: 1e-3, weight decay: 1e-5, dropout: 0.5, epochs: 200.
  \item {\bf SCAN}. $\epsilon=0.5$, $\mu=5$.
  \item {\bf Radar}. Learning rate: 1e-3.
  \item {\bf Anomalous}. Learning rate: 1e-3.
  \item {\bf GCN}. Hidden dimension: 64, batch size: 64, learning rate: 1e-3, weight decay: 0, dropout: 0.5, epochs: 200.
  \item {\bf GraphSAGE}. Hidden dimension: 64, batch size: 64, learning rate: 1e-3, weight decay: 0, dropout: 0.5, epochs: 200.
  \item {\bf GCNAE}. Hidden dimension: 64, batch size: 64 learning rate: 1e-3, weight decay: 1e-5, dropout: 0.5, epochs: 200.
  \item {\bf Dominant}. Hidden dimension: 64, batch size: 64 learning rate: 1e-3, weight decay: 1e-5, dropout: 0.5, epochs: 200.
  \item {\bf DONE}. Hidden dimension: 64, batch size: 64 learning rate: 1e-3, weight decay: 1e-5, dropout: 0.5, epochs: 200.
  \item {\bf ADONE}. Hidden dimension: 64, batch size: 64 learning rate: 1e-3, weight decay: 1e-5, dropout: 0.5, epochs: 200.
  \item {\bf AnomalyDAE}. Hidden dimension: 64, batch size: 64 learning rate: 1e-3, weight decay: 1e-5, dropout: 0.5, epochs: 200, $\alpha=0.5$, $\theta=10$, $\eta=5$.
  \item {\bf GAAN}. Hidden dimension: 64, batch size: 64 learning rate: 1e-3, weight decay: 1e-5, dropout: 0.5, epochs: 200.
  \item {\bf GUIDE}. Feature hidden dimension: 64, structure hidden dimension: 5, batch size: 64 learning rate: 1e-3, weight decay: 1e-5, dropout: 0.5, epochs: 200.
  \item {\bf CONAD}. Hidden dimension: 64, batch size: 64 learning rate: 1e-3, weight decay: 1e-5, dropout: 0.5, epochs: 200.
  \item {\bf DeepSAD}. Batch size: 128, learning rate: 1e-3, epochs: 50.
  \item {\bf REPEN}. Hidden dimension: 20, batch size: 256, epochs: 1000.
  \item {\bf PReNet}. Batch size: 512, learning rate: 1e-3, epochs: 10.
\end{itemize}

\section{Generalization}
We are going to discuss generalization based on four points, 1) the type of applications, 2) the extensibility and reproducibility of the data collection 3) the type of injected anomalies and system anomalies, 4) problems involving DAGs.
\begin{itemize}
  \item \textbf{Type of Applications:} In our submission we presented 3 workflow applications that are described using an open-source workflow management tool and they are representative of the High Throughput Computing (HTC) workloads observed in the community.
        In previous work researchers have used similar workflows to provide a characterization of the scientific workflows seen in the wild~\citep{juve2013characterizing},\citep{krawczuk-works-2021}.
        We selected these workflow applications for the size and complexity of their DAGs (Figure \ref{fig:1000genome}, Figure \ref{fig:montage}, Figure \ref{fig:predict-future-sales}), the ability to acquire real execution traces within a reasonable time for each workflow run and because they are addressing real world problems.
        By using the binaries executed in each node of the workflow we are not claiming that we can transfer learning directly to other applications and execution systems.
        Since each binary can be executing different instructions and can be compiled with different optimizations in mind (e.g., infiniband drivers, gpu accelerators)~\citep{Phillips2020},\citep{kenny2020indis}.
        Additionally, hardware differences among execution environments can lead to different node level performance, overall workflow performance and potentially different results.
        The majority of the workflow management systems model their workflows as DAGs~\citep{ferreiradasilva-fgcs-2017},\citep{mitchell2019btsd}, with dependencies among jobs, creating parent and children relationships, and this dataset captures resource statistics and the behavior of the workflow, tightly coupled with each workflow step.
        Most datasets that currently exist do not offer this information between application and resources, because they are observing the behavior of the system, instead of the behavior of the application, or because of privacy and security policies.

  \item \textbf{Data Collection}: To generate this dataset, we used resources widely available to the educational community.
        We used the Chameleon testbed~\citep{keahey2020lessons} to create our computational infrastructure.
        And we used publicly available workflows described using the Pegasus Workflow Management System~\citep{deelman-cise-2019} and tools that are distributed with it~\citep{papadimitriou2021end}.
        As a result, if one wants to replicate, extend, or incorporate the monitoring and collection into their workflows it’s possible.
        Additionally, we are planning to share our recipes and code for our experimental harness for anybody that might be interested in using it.

  \item \textbf{Injected Anomalies}: In our work we define anomalies as system slowdowns and not hard failures. More specifically with the CPU anomaly we aim to demonstrate a CPU slowdown due to co-location of tasks on the same host, competing for cpu cycles. We introduce this anomaly via cgroups, restricting our binaries to use only a certain number of cores available to the host. For example when a host has 4 cores and we take away 2 cores, we label this anomaly setting as “CPU\_2”, in order to mimic the behavior of tasks that have been mapped to the host and are taking 100\% of the cpu time on the takeaway cores. This can be caused by misconfiguration or even by poor queue priority choices by the user. Similarly with the HDD anomaly we aim to demonstrate filesystem slowdown due to IO overload or filesystem health status (e.g., filesystem repair). We apply the HDD anomaly using cgroup limits and we limit the worker hosts average read and average write speeds. For example, with the HDD\_10 label, we limit the read to 20MB/s and write to 10MB/s. The type of injected anomalies aim to mimic slow downs observed in production systems. For example, at National Energy Research Scientific Computing Center (NERSC) during the execution of the XFEL workflow they discovered a slow down on the shared filesystem, that was delaying or stalling the execution of the tasks~\citep{blaschke2021realtime}.

  \item \textbf{Generalization of DAG datasets}. Not only in the domain of computational workflows but also in other domains, DAGs are widely used to model the dependencies among tasks.
        For instance, DAGs can be used in social network analysis to model the relationships between individuals, organizations, or communities. In this domain, DAGs can help identify influential individuals or clusters, detect communities, and analyze the spread of information or behaviors.
        In biology, DAGs can represent gene regulatory networks, where nodes represent genes and edges represent regulatory interactions between them.
        This allows researchers to study the behavior of genetic systems and identify potential drug targets.
        In traffic management, DAGs can model traffic flow and help optimize traffic signal control, routing, and traffic assignment.
        In finance, DAGs can represent relationships between financial instruments, such as stocks, bonds, and currencies, allowing for portfolio optimization and risk analysis.
        These examples demonstrate the versatility and generalization of DAGs in various domains, making them a valuable tool for understanding and analyzing complex systems.
\end{itemize}

\section{Other Open Datasets}
There are other open datasets that in the past, they have been used to analyze and predict system failures\citep{oliner2007dsn},\citep{Zarza2010_PDPTA},\citep{zheng2011}, and improve scheduling algorithms and cluster health~\citep{wilkes2011more},\citep{tirmazi2020eurosys}. However, these datasets are not suitable for modeling and predicting workflow application performance since they are connections between the workflow application DAGs and the system usage, which we offer in our dataset. A dataset that observes the behavior of the workflow from the application perspective can be found in WfCommons~\citep{coleman2021fgcs}, but there are only 180 traces total that exist in the dataset, without any training labels, making the use of these data insufficient for training ML models.




\begin{figure}[ht]
  \centering
  \subfloat[1000 Genome]{
    \includegraphics[width=0.3\linewidth]{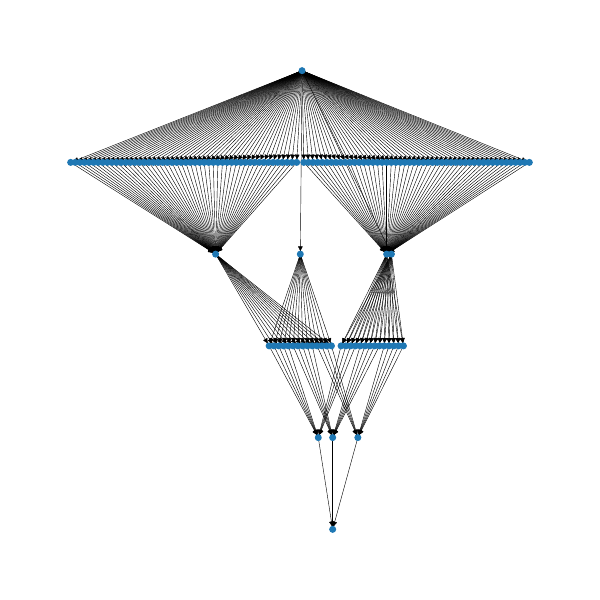}
    \label{fig:1000genome-dag}
  }
  \hfill
  \subfloat[CASA Nowcast]{
    \includegraphics[width=0.3\linewidth]{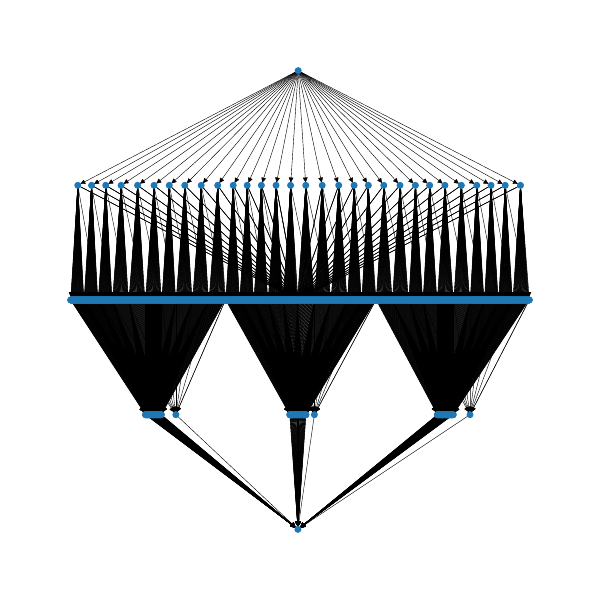}
    \label{fig:casa-nowcast-dag}
  }
  \hfill
  \subfloat[CASA Wind]{
    \includegraphics[width=0.3\linewidth]{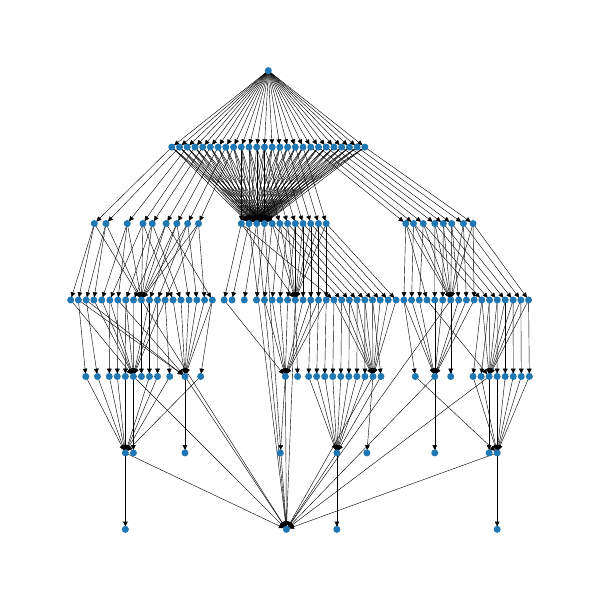}
    \label{fig:casa-wind-dag}
  }
  \\
  \subfloat[EHT Difmap]{
    \includegraphics[width=0.3\linewidth]{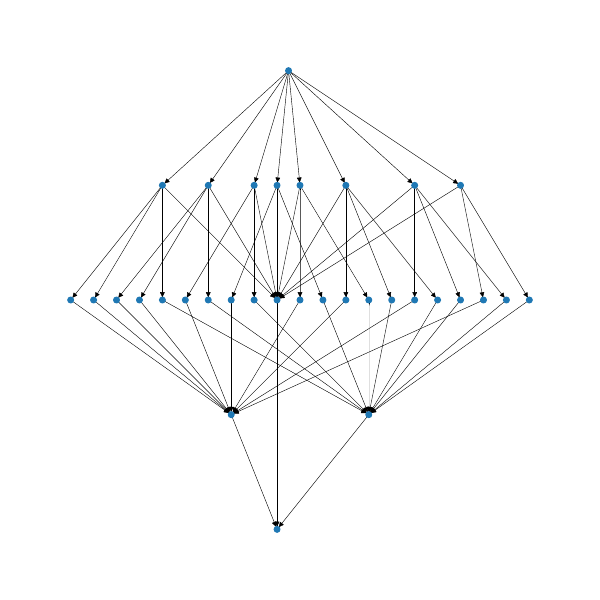}
    \label{fig:eht-difmap-dag}
  }
  \hfill
  \subfloat[EHT Imaging]{
    \includegraphics[width=0.3\linewidth]{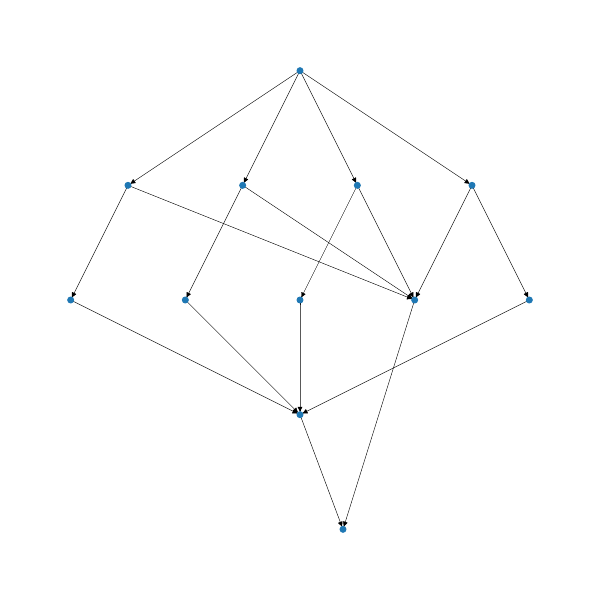}
    \label{fig:eht-imaging-dag}
  }
  \hfill
  \subfloat[EHT Smili]{
    \includegraphics[width=0.3\linewidth]{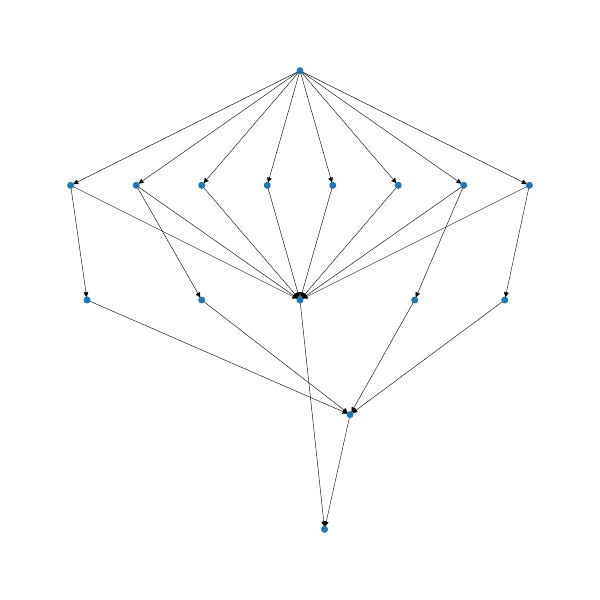}
    \label{fig:eht-smili-dag}
  }
  \\
  \subfloat[Montage]{
    \includegraphics[width=0.3\linewidth]{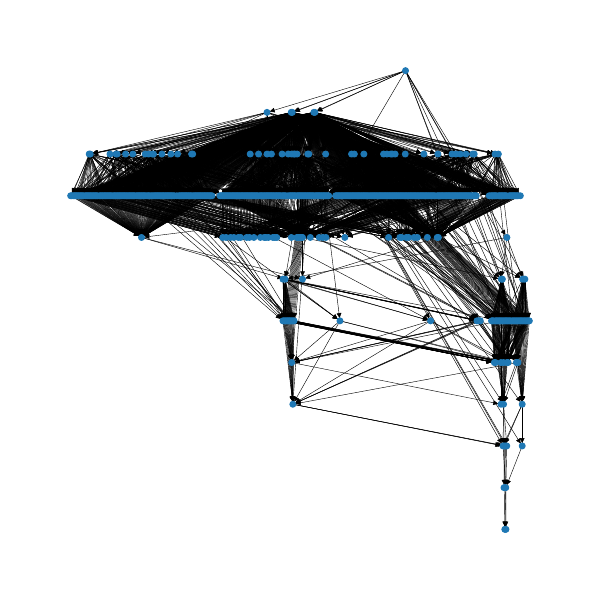}
    \label{fig:montage-dag}
  }
  \hfill
  \subfloat[Predict Future Sales]{
    \includegraphics[width=0.3\linewidth]{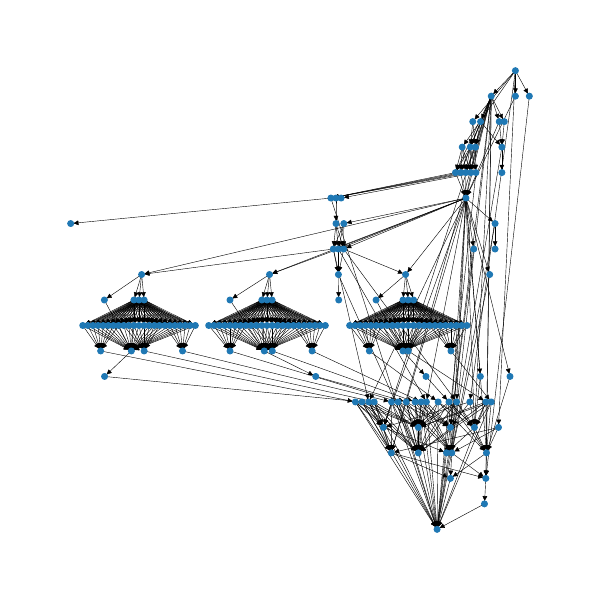}
    \label{fig:future-sales-dag}
  }
  \hfill
  \subfloat[PyCBC Inference]{
    \includegraphics[width=0.3\linewidth]{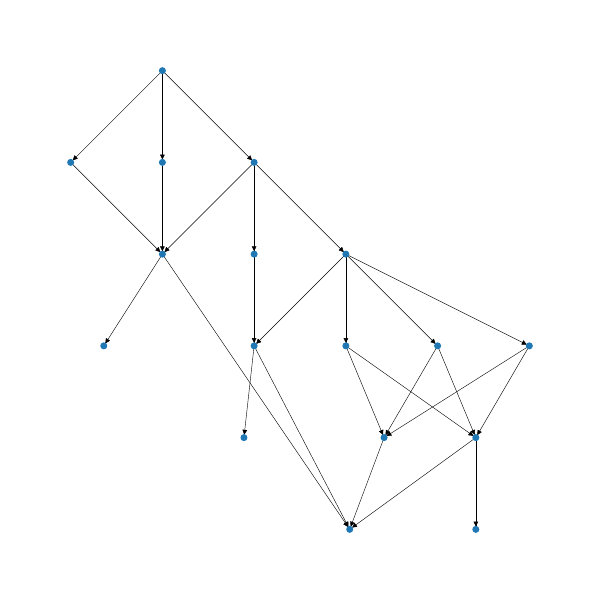}
    \label{fig:pycbc-inference-dag}
  }
  \\
  \subfloat[PyCBC Search]{
    \includegraphics[width=0.3\linewidth]{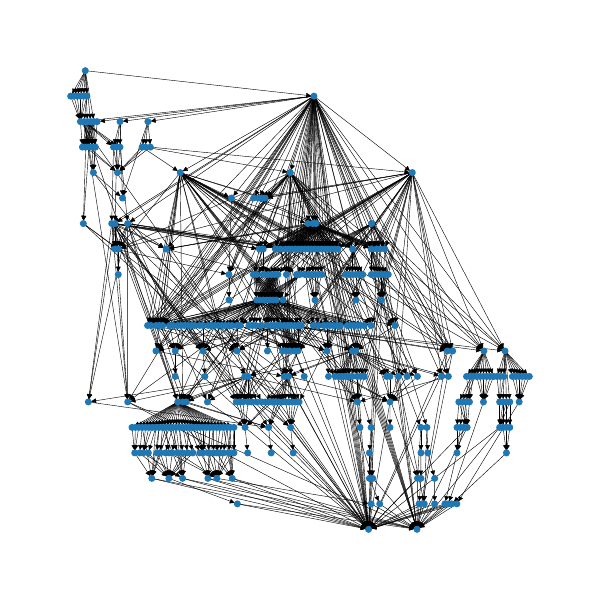}
    \label{fig:pycbc-search-dag}
  }
  \hfill
  \subfloat[Somospie]{
    \includegraphics[width=0.3\linewidth]{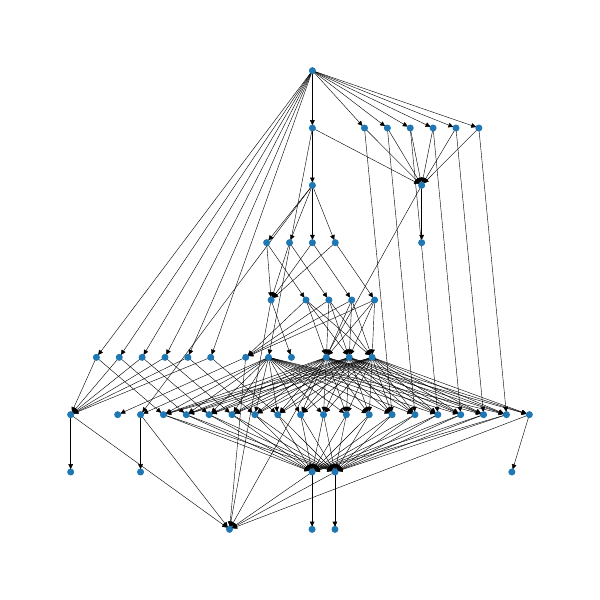}
    \label{fig:somospie-dag}
  }
  \hfill
  \subfloat[Variant Calling]{
    \includegraphics[width=0.3\linewidth]{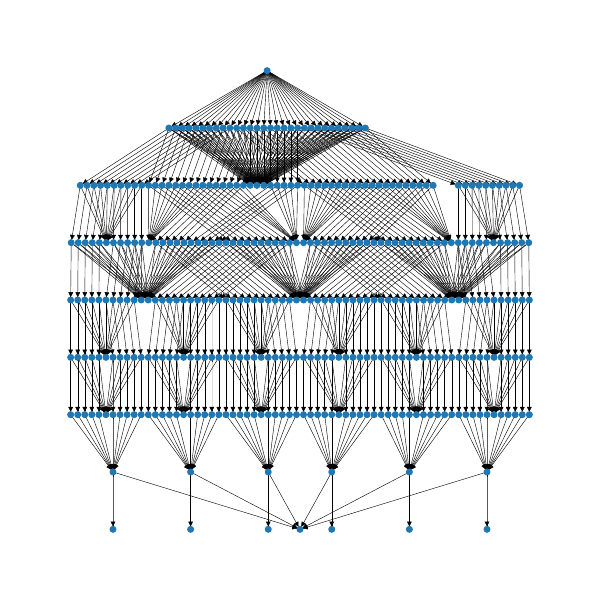}
    \label{fig:variant-calling-dag}
  }
  \caption{Workflow visualizations}
\end{figure}

\section{Supplementary Benchmarks}
\label{appx:benchmark_results}

\begin{figure}[ht]
  \centering
  \includegraphics[width=0.9\textwidth]{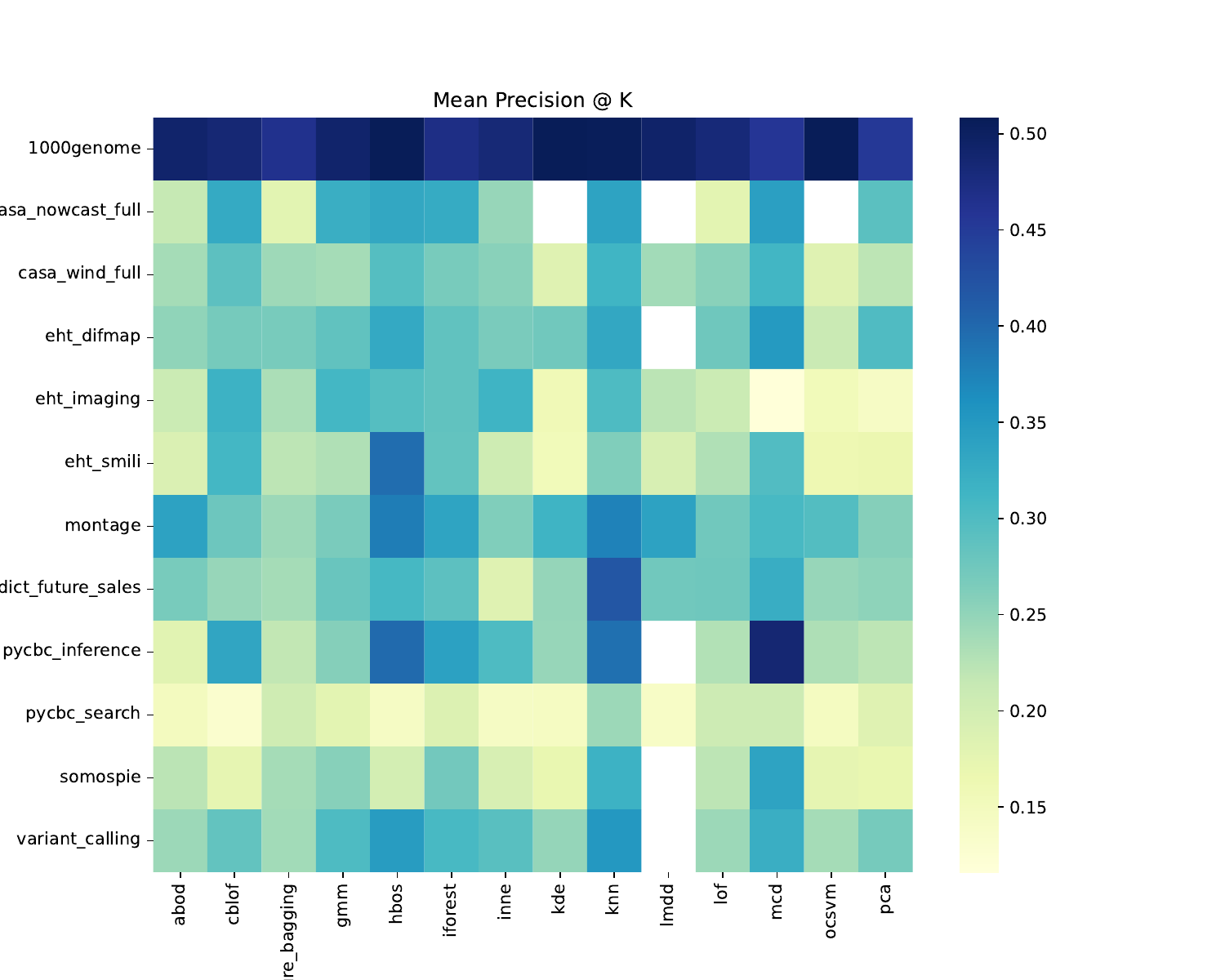}
  \caption{Heatmap of the PyOD algorithms on the benchmark datasets. Reported mean precision score at K.}
  \label{fig:pyod_heatmap}
\end{figure}

We include the complementary benchmark methods here, including semi-supervised ( DeepSAD,  PReNet) methods provided from ADBench~\citep{han2022adbench}.
To expand Table~\ref{tab:ul_results} and \ref{tab:sl_results}, we report the additional results in Table~\ref{tab:model_comp_complementary}, which reports the mean accuracy scores across 5 different runs, even though the unsupervised methods do not require labels for training.
Some key observations that can be made from the table:
\begin{itemize}
  \item The supervised and semi-supervised models generally outperform the unsupervised PyOD and PyGOD models.
  \item The SFT-LLM models (BERT and RoBERTa-large) also perform well, often among the top models for each workflow. Interesting, with a relative model size of 110 M in BERT and 355 M parameters in RoBERTa-large, and parameter efficient training LoRA, is able to achieve competitive performance, showing the training time in Table~\ref{tab:sl_results}.
  \item There are some cases where certain models fail to complete the task, indicated by "TLE" (Time Limit Exceeded) or "OOM" (Out Of Memory). This is due to the complexity of the model processing the graph, especially for the larger workflows and more runs in datasets. Even with advanced hardware accelerators, e.g., NVIDIA A100 in our experiments, the model may still fail to complete the task. This is a common issue in graph-based deep learning models, particularly when processing large graphs and their entire structure.
  \item The performance of the models varies across the different workflows, highlighting the importance of evaluating on multiple real-world datasets.
\end{itemize}
The table provides a comprehensive comparison of the different anomaly detection models, allowing researchers to evaluate their performance across a wide range of datasets and identify the most suitable approaches for their specific use cases.
In addition, due to the most accessible and widely used models from PyOD, we also present the results evaluated with top K precision scores with different models in Figure~\ref{fig:pyod_heatmap}.
It allows to compare the performance of the models across different datasets and identify the most effective models for each workflow.

\begin{table}[th]
  \centering
  \caption{Model Comparison - with selected models from different approaches.}
  \label{tab:model_comp_complementary}
  \resizebox{\columnwidth}{!}{
    \begin{tabular}{l|rrrr|rrrr|rr|rr|rr}
      \toprule
                           & \multicolumn{4}{c|}{PyOD}
                           & \multicolumn{4}{c|}{PyGOD}
                           & \multicolumn{2}{c|}{Supervised}
                           & \multicolumn{2}{c|}{Semi-supervised}
                           & \multicolumn{2}{c}{SFT-LLMs}
      \\
      workflow             & GMM                                  & KNN  & LMDD & PCA  & GAE  & DONE & CONAD & AnomalyDAE & GCN  & MLP  & DeepSAD & PReNet & BERT & RoBERTa-large \\
      \midrule
      1000genome           & .660                                 & .691 & .673 & .626 & .643 & .655 & 0.677 & .664       & .794 & .799 & .708    & .742   & .776 & .808          \\
      casa nowcast full    & .768                                 & .774 & TLE  & .760 & .746 & OOM  & OOM   & OOM        & .784 & .770 & .745    & .729   & .796 & .768          \\
      casa wind full       & .757                                 & .779 & .793 & .749 & .760 & OOM  & 0.749 & OOM        & .791 & .774 & .755    & .766   & .795 & .812          \\
      eht difmap           & .748                                 & .772 & TLE  & .752 & .741 & .772 & OOM   & OOM        & .786 & .775 & .767    & .761   & .784 & .807          \\
      eht imaging          & .815                                 & .794 & .839 & .728 & .728 & .824 & 0.710 & .731       & .817 & .850 & .774    & .796   & .825 & .819          \\
      eht smili            & .768                                 & .776 & .805 & .730 & .742 & .813 & 0.733 & .728       & .805 & .820 & .784    & .768   & .840 & .842          \\
      montage              & .755                                 & .780 & .795 & .748 & .735 & OOM  & 0.745 & .766       & .795 & .828 & .780    & .773   & .803 & .805          \\
      predict future sales & .773                                 & .825 & .807 & .770 & .790 & OOM  & 0.801 & .825       & .814 & .839 & .802    & .826   & .861 & .813          \\
      pycbc inference      & .793                                 & .844 & TLE  & .782 & .776 & OOM  & OOM   & OOM        & .815 & .827 & .819    & .803   & .829 & .808          \\
      pycbc search         & .762                                 & .795 & .759 & .763 & .776 & .795 & 0.783 & .783       & .809 & .800 & .778    & .750   & .825 & .790          \\
      somospie             & .764                                 & .799 & TLE  & .725 & .699 & OOM  & OOM   & OOM        & .801 & .795 & .745    & .740   & .809 & .810          \\
      variant calling      & .762                                 & .773 & TLE  & .751 & .764 & OOM  & OOM   & OOM        & .788 & .819 & .780    & .797   & .791 & .778          \\
      \bottomrule
    \end{tabular}

  }
\end{table}

\end{document}